\documentclass[fleqn,usenatbib,useAMS]{mnras}

\usepackage{graphicx}	
\usepackage{amsmath}	
\usepackage{multicol}        
\usepackage{bm}		
\usepackage{pdflscape}	
\usepackage[T1]{fontenc}
\usepackage{ae,aecompl}
\usepackage{newtxtext,newtxmath}
\usepackage{blindtext}
\usepackage[whole]{bxcjkjatype}
\usepackage[normalem]{ulem}
\usepackage{lineno}
\usepackage{here}
\usepackage{color} 
\usepackage{multirow}
\usepackage{ulem}
\usepackage{float}
\usepackage[title,titletoc]{appendix}
\usepackage{setspace}
\usepackage{booktabs} 
\usepackage{xcolor}
\definecolor{bblue}{rgb}{0.06, 0.06, 0.7}

\renewcommand{\textbf}[1]{{#1}}
\newcommand{\jm}[1]{{}}
\newcommand{\jmold}[1]{{}}

\newcommand{\UVC}{\textit{u-v}~}

\newcommand{\FIMUS}{$50~\mu \rm as$~}
\newcommand{\SGR}{Sgr~A$^{*}$}

\newcommand{\virgoa}{M\hspace{0.1em}87\hspace{0.3mm}}

\newcommand{\DEG}{^{\circ}}

\newcommand{\etal}{et al.}

\newcommand{\RS}{$R_{\rm S}$}

\def\jm#1{{\bf[#1 -- JM]}}

\title{An Independent Hybrid Imaging of \SGR~from the Data in EHT 2017 Observations}
\author[Miyoshi et al.]
{
Makoto Miyoshi$^{1}$\thanks{Contact e-mail:\href{mailto:makoto.miyoshi@nao.ac.jp}{makoto.miyoshi@nao.ac.jp}},
Yoshiaki Kato$^{2}$\thanks{email:\href{mailto:yoshi\_kato@met.kishou.go.jp}{yoshi\_kato@met.kishou.go.jp}},
Junichiro Makino$^{3}$\thanks{e-mail:\href{mailto:makino@mail.jmlab.jp}{makino@mail.jmlab.jp}}
\\
$^{1}$National Astronomical Observatory, Japan, 2-21-1, Osawa, Mitaka, Tokyo, 181-8588, Japan \\
$^{2}$Japan Meteorological Agency: 3-6-9 Toranomon, Minato City, Tokyo 105-8431, Japan \\
$^{3}$Department of Planetology, Kobe University, 1-1 Rokkodaicho, Nada-ku, Kobe, Hyogo 650-0013, Japan
}
\date{Published: 01 May 2024;Accepted 23 April 2024. Received 19 April 2024; in original form 31 October 2023}
\pubyear{2024}

\begin{document}
\label{firstpage}
\pagerange{\pageref{firstpage}--\pageref{lastpage}}
\maketitle
\begin{abstract}
We propose that the ring structure found by the Event Horizon Telescope Collaboration (EHTC) as the black hole shadow of \SGR~is an artifact by the bumpy PSF~(Point Spread Function) of the EHT2017. The imaging using sparse \UVC data requires detailed scrutiny of the PSF.
The estimated shadow diameter~($\mathrm{48.7\pm7~\mu \rm as}$) is equal to the spacing between the main beam and the first sidelobe of the PSF~($\mathrm{49.09~\mu \rm as}$), which immediately suggests a potential problem in the deconvolution of the PSF. We show that the ring image can be derived from non-ring simulated datasets~(noise only; point source) with a narrow Field-of-View~(FOV) and an assumed self-calibration suggesting the EHT2017's \UVC coverage is insufficient for reliable imaging.
The EHTC analysis, based on calibrations with assumptions about the source's size and properties, selected the final image by prioritizing  appearance rate of the similar structure from a large imaging parameter space over data consistency.
Our independent analysis with the conventional hybrid mapping reveals an elongated east-west structure, consistent with previous observations. We believe it to be more reliable than the EHTC image, due to half the residuals in normalized visibility amplitude.
 The eastern half is brighter, possibly due to a Doppler boost from the rapid rotating disk.
We hypothesize our image shows a portion of the accretion disk about 2 to a few \RS ~away from the black hole, rotating with nearly 
$\mathrm{60~\%}$ of the speed of light viewed from an angle of $\mathrm{40-45\DEG}$.
\end{abstract}
\section{Introduction}\label{Sec:Intro}
\subsection{Study Overview}\label{Sec:Intro-1}
~Our independent analysis of the EHT 2017 observational data of \SGR~, the same dataset used by the EHTC, shows an east-west elongated shape~(Figure~\ref{Fig:ourimages}), consistent with previous millimeter wave observations, however, distinctly different from the EHTC's ring-like structure.
The structure we obtained is more reliable than the EHTC ring image, on the basis that the residuals in the normalized visibility amplitude that 
half the size of those in the EHTC ring image~(Section~\ref{Sec:NAR}) though the residuals in closure quantities show comparable levels~(Sections~\ref{Sec:RCP},\ref{Sec:RCA}). 
We hypothesize that our image indicates 
that the black hole shadow of \SGR reported by the EHTC
is an artifact caused by imperfect deconvolution of the bumpy PSF~(Point Spread Function) of the EHT 2017 observation~(Figure~\ref{Fig:psf47}). 
This dataset is likely to produce spurious $50~\mu \rm as$ interval structures in the imaging results.
The PSF structure has the first sidelobes with a height comparable in intensity to the main beam~($\mathrm{\sim49~\%}$ level), separated from the main beam by $\mathrm{49.09~\mu \rm as}$.
There is a very deep negative minimum (~$\mathrm{-89~\%}$ level of the main beam) at the midpoint between them. 
We found that this data sampling can plausibly produce a ring structure with a diameter of $\sim50~\mu \rm as$, even in the case where the data set contains point or noise information.\\
~The EHTC estimated shadow diameter~is exactly the same as the spacing between the main beam and the first sidelobes in the PSF.
The sub-structures in the EHTC ring are also similar to the PSF structure.
The configuration of the three prominent bright spots on the EHTC ring is comparable to the positions of the main beam, the northern first sidelobe, and the eastern sidelobe in the PSF structure~(Figure~\ref{Fig:psf47}).
The shadow in the center of the EHTC ring has the same shape and size as the default restoring beam obtained by Gaussian fitting to the main beam in the PSF, even though the EHTC image is produced with a $20~\mu \rm as$ circular restoring beam~(Figure~\ref{fig:default-beam}).
The similarities between the EHTC ring and PSF structures described above suggest potential problems in the deconvolution of the PSF during the EHTC imaging process.
In the analysis of the EHTC data, the amplitude calibration is performed with an observed source size assumption of $\mathrm{60~\mu \rm as}$, though partially applied. 
To mitigate the time variability in the \SGR structure, data weighting strategies derived from general relativistic magnetohydrodynamical~(GRMHD) simulations with the source size assumption are used.
In addition, the EHTC analysis used a unique criterion for selecting the final image. It is not based on consistency with the observed data, but rather on the highest rate of appearance of the image morphology
from a wide imaging parameter space.
Our concern is that these methods mentioned above may interfere with the PSF deconvolution and cause the resulting image to reflect more structural features of the PSF than the actual intrinsic source structure.

On the other hand, our independent analysis used the traditional VLBI imaging techniques to derive our final image. We used the hybrid mapping method, which is widely accepted as the standard approach; iterations between imaging by the CLEAN algorithm and calibrating the data by self-calibration, following well-established precautions. In addition, we performed a comparison with the PSF structure, noting the absence of distinct PSF features in the resulting images. Finally, we selected the image with the highest degree of consistency with the observational data. 

Our final image shows that the eastern half of this structure is brighter, which may be due to a Doppler boost from the rapid rotation of the accretion disk. We hypothesize that our image indicates that a portion of the accretion disk, located approximately 2 to several \RS~away from the black hole, is rotating at nearly $\mathrm{60~\%}$ of the speed of light and is seen at an angle of 
$\mathrm{40-45\DEG}$~(Section~\ref{Sec:results}).

In this paper we report the results of our analysis of the public EHT data of the \SGR.
Immediately following this, as part of Introduction, an overview of previous \SGR studies based on radio observations~(Section~\ref{Sec:Intro-2}) and a summary on EHTC2022 papers~(Section~\ref{Sec:Intro-3}) are given.
We describe the observational data released by the EHTC in Section~\ref{Sec:obs+data}, our data calibration and imaging process in Section~\ref{Sec:ourreduction} and Appendix~\ref{Sec:snplt}, and our imaging results in Section~\ref{Sec:results}.
We then examine how the EHTC ring of \SGR~was found
in Section~\ref{Sec:the EHTC-ring}.
In Section~\ref{Sec:DISC}, we discuss our results and the future black hole imaging, and our conclusion is given in Section~\ref{Sec:CR}.
In Appendix~\ref{Sec:uv46},
we show that the \UVC coverage and the corresponding PSF structure of the April 6, 2017 data. We explain the difference from those of the April 7, 2017 data and show why the EHTC met the more difficulty to find the $\mathrm{50~\mu \rm as}$~ring shape from the April 6 data. 
In Appendix~\ref{Sec:betweeIFs}, we show the differences in closure phase and amplitude between the two simultaneously recorded channels of all EHTC public data of \SGR~observations.
There seems to be inconsistency in the closure amounts between channels, and there is a possibility that the observed source information is not being accurately stored in the data.

\subsection{Prior radio research of \SGR}\label{Sec:Intro-2}
\SGR~is a supermassive black hole~(SMBH) at the center of our Milky Way galaxy, first detected through radio observations.
It was discovered in 1974 by the NRAO radio interferometer (S-X bands) as a compact radio source in the Sgr~A~region, a complex radio source at the Galactic center.
Constant visibility amplitude indicated that its apparent size was less than \textnormal{0.1~arcsec} (corresponding to a real size of \textnormal{800~AU}), and its observed brightness temperature exceeded $\mathrm{10^7~\rm K}$, suggesting that it is a black hole~\citep{Balick+Brown:74}.

Since 1990s, advances in infrared observational techniques have enabled the observation of star motions around \SGR~, from which the mass of \SGR~has been accurately measured. 
The high mass density of its inner volume has led scientists to conclude that it is a black hole~\citep{Eckart:96,Ghez:00,Munyaneza:02,Schodel:02}. 
The distance from the Earth to the Galactic center is about \textnormal{8~kpc}, which is orders of magnitude closer than the distances to supermassive black holes in other galaxies. 
Considering the mass of the \SGR~black hole ($\mathrm{\sim 4\times 10^6M_{\odot}}$), the angular size of the Schwarzschild radius~(\RS) viewed from the Earth is about $\mathrm{10~\mu \rm as}$, making \SGR~the most important object for the study of the structure near a black hole~\citep{Falcke:00,Miyoshi:04,Miyoshi:07}.

However, despite the high spatial resolution of ~Very Long Baseline Interferometer~(VLBI), the measurements were only of the apparent size of the \SGR~image~\citep{Doeleman:01, Bower:98, Yusef-Zadeh:94, Rogers:94, Alberdi:93, Krichbaum:93, Marcaide:92, Lo:85, Lo:98}. 
This is due to the scattering of the radio image of \SGR~by the intervening plasma, causing the observed image size to increase as the square of the observed wavelength~\citep{Davies:1976,vanLangevelde:92,Bower:04,Shen:05}. 
To observe the intrinsic structure of \SGR  free from the effects of scattering, higher frequency observations at millimeter to submillimeter wavelengths are needed.

 In addition, there was another difficulty related to data calibration. The Very Long Baseline Array (VLBA), which was expected to be a full-scale VLBI instrument, but whose stations were all located in the northern hemisphere, had to be viewed from a low elevation angle in order to observe the southern sources including \SGR. 
Such observations are subject to phase fluctuations due to atmospheric water vapor, and especially at millimeter wavelengths, phase calibrations were very difficult.
Also the VLBA's \UVC coverage becomes sparse for \SGR~located at $\mathrm{\delta \sim-30~\DEG}$, making it difficult to obtain good images (\cite{Zensus:99},~and papers in \cite{GCW:99}).

Using the closure phase and amplitude, which are independent of antenna-based observational errors, the \SGR~images at \textnormal{43}~and \textnormal{86~GHz} were measured.
These measurements in visibility domain revealed an ellipse elongated mainly in the east-west direction~\citep{Bower:04,Shen:05}. 
However, these results were not yet considered to be a measurement of the intrinsic figure, but of the structure of the \SGR~image, which is still broadened by scattering.
Later, however, \cite{Bower:2014} found by analyzing the closure amplitudes of the \textnormal{43~GHz} VLBA data with increased sensitivity that the intrinsic image of~\SGR~itself extends much further in the east-west direction than in the north-south direction. The intrinsic source is modeled as an elliptical Gaussian with a major axis of $\mathrm{354\times126~\mu \rm as}$ and a position angle of~$\mathrm{95\DEG}$. 
This means that the scattered image is indeed strongly elongated in the east-west direction due to the anisotropy of the scattering, but the original image of the \SGR~itself is also elongated in the east-west direction.

In these previous studies, 
the shape of \SGR~was determined by fitting a model to visibility data that is assumed a point-symmetric structure, such as elliptical Gaussian shapes, without considering asymmetric components. 
However, some of the CLEAN maps in those studies deviate slightly from the point-symmetric structure. 
For example, the super-resolution image by CLEAN shown in Figure 1b of~\cite{Shen:05} has an elongated shape in the east-west direction and is asymmetric between the eastern and western halves.

The next step was to search for deviations from point-symmetric structures of \SGR.
In an attempt to analyze the periodic short-term variations of \SGR~in image domain, 
\cite{Miyoshi:11} used the Slit-Modulation Imaging (SMI) method~\citep{Miyoshi:08}, which is free from the influence of differences in \UVC coverage between snapshot maps.
Although the detection reliability is not high, they found non-zero closure phases in the data from \textnormal{43~GHz} observations of \SGR~taken after its near-infrared flare event. 
A single static image did not fully explain the data, indicating the presence of short-term variations in the structure of \SGR.
Furthermore, they found that the structure of \SGR~is elongated in the east-west direction with asymmetry and that the time variation spectra show shorter periodic variations on the east side than on the west side. 
Furthermore, an asymmetry in its east-west elongation has been reported based on subsequent observations and analyses.
To further investigate asymmetric components of the \SGR~structure, large antennas such as the Robert C. Byrd Green Bank Telescope (GBT) and the Large Millimeter Telescope Alfonso Serrano (LMT) were included in the VLBA to increase sensitivity. \cite{Brinkerink2016} investigated the 86~GHz VLBA visibility data using closure phase analysis and found evidence of an eastern secondary source component located approximately $\mathrm{\sim100~\mu \rm as}$ from the primary component of \SGR~, and also pointed that other results by 
\cite{Fish:16} at \textnormal{230~GHz}, 
\cite{Ortiz-Leon:16} at \textnormal{86~GHz},
and 
\cite{Rauch:16} at \textnormal{43~GHz} 
indicate asymmetric emission of \SGR~at different frequencies and over different time periods. 
Thus, the east-west elongation and its asymmetry of \SGR~have been reported from such as VLBA observations, which provide lower frequency observations than EHT 2017, but have much more \UVC coverage than EHT 2017.

\cite{Johnson:2018} show that the angular broadening has
an FWHM of
$\mathrm{(1.380 \pm 0.013)\times \lambda_{cm}^2~\rm mas}$ along the major axis 
and 
$\mathrm{(0.703 \pm 0.013)\times \lambda_{cm}^2~\rm mas}$ along the minor axis, 
with the major axis at a position angle $\mathrm{81^{\circ}.9 \pm 0^{\circ}.2}$.
\cite{Johnson:2018} also estimated the intrinsic size of the \SGR~to be proportional to the wavelength between \textnormal{1.3~mm}~(\textnormal{230~GHz}) and \textnormal{1.3~cm}~(\textnormal{22~GHz}), that is, approximately $\mathrm{\theta_{\mathrm{src}} \sim 0.4~\times \lambda_{cm}~\rm mas}$.
It follows that at \textnormal{43~GHz} the angular broadening is larger than the intrinsic size of \SGR~; at \textnormal{86~GHz} the two are expected to be comparable, and at \textnormal{230~GHz} the intrinsic size of \SGR~is expected to dominate~(Table~\ref{tab:scattering}).\\
\begin{table}
\centering
\begin{tabular}{@{}rrrr@{}}
\toprule
Frequency~~~~~~~~~~~~~~~~~~~~~~~~& 43~  & 86~  & 230~~(\rm GHz)\\
\midrule
Broadening~(Major~axis)& 67.2& 16.8& 2.3~~(\RS)\\
~~~~~~~~~~~(Minor~axis)& 34.2&  8.6& 1.2~~(\RS)\\
~~~~~~~(Position~Angle)& \multicolumn{3}{c}{$81^{\circ}.9 \pm 0^{\circ}.2$} \\ \midrule
Intrinsic~Size~~~~~~~~~~~~~~~~~~~~& 27.9& 14.0& 5.2~~(\RS) \\
\bottomrule
\end{tabular}
\caption{
Angular broadening of the image of Sgr~A$^{*}$ and an estimated intrinsic size, based on calculations following~\citep{Johnson:2018} assuming $1~R_{S}~=~10~\mu \rm as$.}
\label{tab:scattering}
\end{table}
~We summarize what was known about the image of~\SGR~prior to the full-scale observation of the EHT in 2017.
\vspace{-2mm}
\begin{enumerate}
\item
The apparent image broadens as the square of the observed wavelength mainly due to scattering. The scattering is anisotropic. The broadening effect is stronger in the east-west direction. Its effect is expected to be nearly negligible for observations at \textnormal{230~GHz}~\citep{Johnson:2018}.
\item The intrinsic structure of \SGR~at \textnormal{43~GHz} is estimated to be elongated in an east-west direction~\citep{Bower:2014}.
\item
The east-west anisotropy in the intrinsic structure of ~\SGR~has also suggested by some observations at \textnormal{43, 86, \&~230~GHz}.
\end{enumerate}

\subsection{Ring-like image of \SGR~by the EHTC 2022}\label{Sec:Intro-3}
In May 2022, the EHTC reported a ring-like image from the EHT observations of \SGR~, along with the confirmation that \SGR~shows short timescale changes with an intra-hour variability
~\citep{EHTC2022a,EHTC2022b,EHTC2022c,EHTC2022d,EHTC2022e,EHTC2022f}. 
The time variation of the observed source made normal interferometric synthesis imaging difficult, and the EHTC used their original analysis to determine the source structure while addressing for the variability.
The structure reported by EHTC is dominated by a bright, thick ring with a diameter of $\mathrm{51.8 \pm 2.3~\mu \rm as}$. 
The ring exhibits an azimuthal brightness asymmetry, with three bright spots located at $\mathrm{PA\sim-140\DEG,-40\DEG}$, and $mathrm{+70\DEG}$. 
The central hole region was not completely dark, but comparatively dim~\citep{EHTC2022a}.
This shape is reminiscent of their previous \virgoa ring images~\citep{EHTC1,EHTC2,EHTC3,EHTC4,EHTC5,EHTC6}, as if it were a face-on rather than an edge-on view of the black hole shadow and accretion disk.

As for the time variation of \SGR~, its first radio detection was done by \cite{Miyazaki:04}.
They found that \SGR~exhibits intraday variations, or short bursts, in the millimeter wavelength range.
Similar short bursts have also been detected in the X-ray~\citep{Baganoff:01} and near-infrared~\citep{Genzel:03}~regions, strengthening the case for \SGR~as a variable source.
In addition, recent Atacama Large Millimeter/Submillimeter Array~(ALMA) 
observations have clearly shown that \SGR~exhibits intrahour variations in intensity at millimeter and submillimeter wavelengths~\citep{Iwata2020, Miyoshi2019}.
The presence of short-term variations on \SGR~poses a new challenge for current millimeter VLBI techniques to produce accurate images of the source structure, 
as they require observing times of up to 24 hours to achieve full \UVC coverage. 
The fundamental assumption for synthesizing an image using a radio interferometer is that the observed source's brightness distribution remains constant throughout the observations.
In the past, VLBI observations of SS~433 also encountered intensity variations of the observing source during the observation, raising concerns about the impact on image synthesis~\citep{Vermeulen:93}, but the time variations of \SGR~at millimeter wavelengths are much more intense and have a shorter timescale.
Moreover, the variability of \SGR~complicates large-scale imaging of the Galactic Center region with ALMA, resulting in distortions in the synthesized images~\citep{Tsuboi:22}.

We felt a need for an independent analysis of the data to determine the reliability of the ring shape reported by EHTC, which is not consistent with the previously recognized structure of \SGR~, although it is certain that the EHTC ring size is the same as the shadow size expected from relativity.

\section{Observational data}\label{Sec:obs+data}
The EHT observed \SGR~in early April 2017 with a total of eight stations, including the Antarctic station (South Pole Telescope, SPT). 
The inclusion of the SPT station improved the spatial resolution in the north-south direction.

The observations of \SGR~were conducted on five nights between April 5 and 11, 2017~\citep{EHTC2022a}. The EHTC publicly released data from two observations on April 6 and 7. A total of 28 datasets are available to the public, which were calibrated using different methods as described below,
 resulting in different data sets. 
The explanation for these datasets is available in the "README.md" file, which can be accessed at \url{https://datacommons.cyverse.org/browse/iplant/home/shared/commons_repo/curated/EHTC_FirstSgrAResults_May2022}.

Given the extensive data, one might find it challenging to discern patterns; however, several key considerations should be noted.
First, there are two different data sets for the same observation because two different calibration methods were used: EHT-HOPS and rPICARD (CASA) pipelines. Second, \SGR~exhibits short-term intensity variations, and probably structure changes, during the observations. 
To correct for this variability, EHTC made "normalized" data sets 
where each of visibility amplitudes is normalized by the total flux density from the \SGR~'s light curve (observed by ALMA and SMA) at the corresponding time~\citep{Wielgus:22}.
There are additional modified data sets that have been corrected for the LMT station.
To account for the uncertainty in the gain of the LMT station, the amplitude calibration was performed assuming the source size.
According to Section 2-2 of \cite{EHTC2022c}, the LMT amplitude gains were pre-corrected assuming a $\mathrm{60~\mu \rm as}$ source size, as constrained by the baselines shorter than $\mathrm{2~\rm G \lambda}$ (specifically, the SMT-LMT baseline only).
The effectiveness of these corrections in providing accurate calibration remains uncertain. Moreover, there is no concrete evidence to suggest that these corrections preserve the intrinsic information of the observed source structure.

Finally, to obtain short-time-scale dynamic properties of \SGR~, the EHTC sliced the data set for 
\textnormal{100-minute} interval snapshot \UVC-coverages
~(denoted as 'BEST').
Using amplitude corrected data sets with the addition of their own original corrections, EHTC obtained the static image of \SGR~(see Figure 3 in~\cite{EHTC2022a}). 
Our analysis in this paper is based on the data set used by the EHTC to obtain their static image of \SGR. 
This data set was normalized by EHTC using
the total flux density from the \SGR~'s light curve as mentioned.
In the case that the observed object varies uniformly in brightness without changing its overall structure, this normalization of the data will accurately reflect the source structure.
Our data analysis focused on the April 7 data, which the EHTC analyzed in detail~\citep{EHTC2022c}. We selected the HOP-calibrated datasets because the CASA calibration data show a large difference in closure amplitude between channels of the raw data~(see the Table~\ref{Tab:IF-dif-CA}~in Appendix ~\ref{Sec:betweeIFs}).
Namely we analyzed data from April 7, i.e., calibrated by HOP, corrected for amplitude with respect to LMT, and normalized for time variability using measured intensities. 
Our initial self-calibration solution for the phase with a point model yielded a phase solution with a mean close to zero~ (Figure~\ref{Fig:SN1} in Appendix~\ref{Sec:snplt}). According to the README.md, complex calibrations were performed after the correlation process. 
In \citet{EHTC2022b} it is not explicitly stated that a phase self-calibration with a point model was performed in the calibrations of the HOP data, while for the CASA data it is stated that a phase self-calibration with a point model was performed. It is reasonable to infer that all EHT 2017 public data of \SGR~have been calibrated by the one corresponding to the self-calibration with a point model.

Note that our independent analysis cannot follow and fully validate all the calibrations performed by the EHTC. Because the public data are compressed into single sub-channel in each IF channel, we cannot perform fringe searches using tools like FRING or other tasks in AIPS, meaning we are unable to independently search for delay, delay-rate, and gain errors. 
The same is true for bandpass calibration.
 The only calibration method we can perform is phase and amplitude correction via self-calibration~\citep{RefHM3,Cornwell:1981}.

\section{Our data calibration and Imaging}\label{Sec:ourreduction} 
Our data analysis is based on hybrid mapping, which is the most commonly used method for VLBI image analysis.
First, a review of VLBI imaging is presented, followed by the description of our data analysis.
\subsection{VLBI imaging}\label{Sec:Vimaging}
We provide an overview of VLBI imaging techniques, including not only the principles, but also some of the issues that need to be considered in the actual analysis.
\subsubsection{Fundamentals of Synthesis Imaging}\label{Sec:whatisimaging} 
The principle of imaging using radio interferometers, including VLBI (Very Long Baseline Interferometry), is not unique to radio wavelengths but is common to that of optical/infrared telescopes. The resulting image is the convolution of the true brightness distribution of the observed source with the point spread function (PSF) of the radio interferometer. The PSF is commonly referred to as the "dirty beam" in radio interferometry . Also, the resulting image is called the "dirty map". In actual data, the observational errors are added in the convolution of the two. To obtain the true structure of the observed source, it is necessary to deconvolve the PSF while removing the errors.

The PSF or "dirty beam" is determined by the \UVC coverage of the radio interferometer. 
\UVC is a technical term used in radio interferometry and means the spatial Fourier component.
The \UVC coverage and the PSF are Fourier duals of each other and are mathematically equivalent. 
Despite the mathematical equivalence, it is almost impossible to understand the characteristics of a PSF simply by looking at the \UVC coverage plots, so it is not the \UVC plot but the structure of the PSF itself that needs to be presented in a scientific paper.

To be sure, if the PSF is mostly a dominant lobe, there is no need to worry about the negative impact of the PSF structure.
However, if the interferometer consists of only a few element antennas - 
as was often the case in the early days of VLBI (although much less so nowadays)- the \UVC coverage will be sparse and the corresponding PSF structure will be far from a single-point structure.
Therefore, the obtained image (dirty map) will be very different from the true brightness distribution of the observed source, and it is necessary to deconvolve the PSF from the dirty map in order to know the source structure.
There is a risk that the actual PSF deconvolution process will not produce sufficient results and will produce artifact structures derived from the PSF structure.
It is therefore very important to examine the PSF structure, because knowing the structure allows us to predict possible artifacts. (For this reason, the AIPS tasks of image synthesis automatically generate the PSF along with the deconvolved image for convenience.)

\subsubsection{CLEAN algorithm\label{Sec:clean}}
The CLEAN algorithm~\citep{Clark1980,Hogbom1974}, was developed to perform deconvolution of the PSF structure and has been the most widely used in the imaging analysis of radio interferometers. 

The CLEAN algorithm is as follows:
Assume that the observed source structure consists of multiple point sources; examine the dirty map, assume that there is a point at the location of its maximum peak, and remove the spatial Fourier component corresponding to that point from the observed visibility. The spatial Fourier component corresponding to the point source is removed from the observed visibility, not all of the peak intensity, but at most a few percent of the peak intensity. 
The next dirty map is formed by Fourier transforming of  the remaining visibility, then finding the new maximum peak again, and removing that peak as well. 
This iteration is repeated until the intensity of the remaining map reaches the noise level. 
%
The set of points (CLEAN components) obtained from the CLEAN iterative subtractions may be convolved with a somewhat sharper restoring beam.
Note this is not an intrinsic approach for CLEAN.
The CLEAN algorithm does not specify the spatial resolution of visibility data.
Therefore, it is necessary to determine the spatial resolution separately.
A Gaussian shape obtained by fitting  to the main beam of the PSF is typically used as the restoring beam. 
(A narrower beam shape will also be used as the restoring beam if a higher spatial resolution than the main beam size of the PSF can be achieved. In such cases, the map is called a super-resolution map.)
To reflect the actual noise, the last residual map is added to the convolved map to complete the CLEAN map. 
However, its deconvolution performance is not perfect.
The resulting CLEAN map is not completely free from the influence of the PSF structure.
To date, there is no algorithm exists that can achieve complete deconvolution of the PSF structure. 
This limitation also applies to other methods used by the EHTC (Event Horizon Telescope Collaboration), as described in Section~\ref{PSF-in-simulation}.
There are at least two reasons for incomplete deconvolution.
First, the PSF structure contains multiple peaks (sidelobes) in addition to the main beam.
As a result, multiple corresponding peaks appear in the dirty map.
Peaks due to sidelobes may be selected as CLEAN components by mistake.
The PSF structure also has negative minima that create false (negative) peaks during CLEAN iterations. 
This is because the normal clean procedure selects the maximum peak by the absolute value of the pixel.
Especially in the case of sparse \UVC coverage, it is important to be aware of the possibility of obtaining false CLEAN components.
Second, poor data calibration will cause the amplitude and phase of the spatial Fourier components to deviate from those of the structure of the observed source, and then the actual dirty map will deviate from the convolution structure of the PSF and the observed source structure. Again, accurate deconvolution becomes difficult.

\subsubsection{Hybrid mapping process\label{Sec:hmp}}
Calibrating the data is essential in the imaging process to obtain as much of the correct source structure as possible.
In VLBI imaging, relying solely on a priori calibration (based on antenna performance and receiver temperature data) is typically insufficient. 
This necessity has led to the development of the hybrid mapping method, which involves iterative processes between self-calibration and deconvolution algorithm (e.g. CLEAN). Self-calibration is a technique that uses an assumed model image to determine the calibration parameters of each station~\citep{Readhead78,RefHM1,RefHM3}.

During the hybrid mapping process, it is common to start with a point source as the model image for the initial self-calibration. If the nearly exact structure of the observed source is known, it can be adopted as the initial model. However, care  must be taken to ensure that the "nearly exact structure" is objectively established and not a misconception.
When deriving calibration solutions in hybrid mapping, it is safer to estimate only the phase solutions in all steps except the last one. Only in the last step can both phase and amplitude be confidently determined as the final calibration solutions. 
Estimating both phase and amplitude solutions in the early stages of iterations, when the model image is far from the observed source structure, is highly unreliable, especially in the case of sparse \UVC coverage data.
We can see the case in~\cite{Miyoshi:03}.
The jet-like structure of \SGR~reported in the paper, but is an artifact resulting from errors in the repeated model image estimation during the hybrid mapping process. In particular, the frequent use of self-calibration in A\&P mode~(the mode to estimate both phase and amplitude calibration solutions) amplified the image intensity and produced a false jet-like artifact image.

Self-calibration produces a calibration solution that follows the assumed image model. As a result, the image from the "calibrated" data may well resemble the assumed model image rather than the real structure of the observed source. 
This tendency is particularly strong in the case of sparse \UVC coverage
(For example, see Figure~\ref{M87simring} and Section~\ref{ringsim0}).
The use of self-calibration must be done very carefully.

\subsubsection{BOXing technique}\label{BOX}
BOXing, a technique that restricts the field of view, is often used in the deconvolution process~(CLEAN) as a partial solution for the problems of insufficient calibration and poor \UVC coverage. 
However, if the BOX setting does not cover the range of the correct brightness distribution, the deconvolution of the PSF cannot be performed properly and an artifact image will appear~(For example, see Figure~\ref{M87simring} and Section~\ref{ringsim0}).
A BOX placed at a position where there is no emission produces an artifact emission at that position.\\

Once an image is obtained that seemingly appears good, the next step is to objectively confirm its validity.
\subsubsection{Comparison with the PSF structure}\label{PSFchk}
The first step is to check the sign of poor deconvolution of the PSF by comparing the obtained image and the PSF structure. 
In particular, it is necessary to check if there is an interval of bright peaks equal to an integer multiple of the interval between the main beam and the first sidelobe.
If this is the case, it is likely that the source structure has not been correctly captured due to the sidelobes of the PSF.

\subsubsection{Check for consistency with visibility}\label{vischk}
The second step is the check of the consistency between the image and the observed visibility data. 
In principle, there are an infinite number of images that satisfy the set of spatial Fourier components (visibilities) that are obtained with finite \UVC coverage~\citep{Bracewell:54}. 
However, in actual interferometric imaging analysis, obtaining an image that fully satisfies the visibility set proves challenging. The optimal approach is to acquire an image that closely matches the visibility set. The quality of various images derived from the analyses can be evaluated based on their degree of consistency with the visibility set.

The image is inverse Fourier transformed and brought into visibility space, and its phase and amplitude are compared to the correspondences of the observed visibility. For amplitude comparison, it is better to evaluate the normalized amplitude~(Appendix~\ref{Sec:NAR-def}) rather than the raw amplitude.

Here, the observed data are calibrated using the self-calibration solutions obtained with the image as a model, and the calibrated data set is used for the comparison.
In VLBI imaging, obtaining a good image requires sufficient calibration of the data. Comparing the observed visibility without  calibration is likely to yield a very large discrepancy between them, even if the obtained image is correct. Therefore, we need to compare the image with the calibrated observational data.
In other words, this is not an investigation of whether the obtained image is consistent with the observed source structure, but rather an investigation of the degree of self-consistency of the calibration and the image synthesis performed.

\subsubsection{Check for consistency with closure quantity}\label{visclo}
One might then think that there is no method to investigate the degree of consistency with the observed source structure. While somewhat indirect, such a method does exist; the use of the closure phase and closure amplitude quantities.
Closure phases (\cite{Jennison:1958};~Appendix~\ref{Sec:RCP-def}) 
and 
closure amplitudes (\cite{Felli:89,Readhead:80,Twiss:60};~Appendix~\ref{Sec:RCA-def}) are quantities not affected by antenna-based errors and are determined solely by the structure of the observed source.
The closure investigation is crucial to evaluate the obtained image. 
However, we should keep the following three limitations in mind.
\vspace{-0.75em}
\begin{enumerate}
\item
The number of images that satisfy the closure quantity is, in principle, infinite. In the actual analysis, even that one is hard to reach.
\item
Although antenna-based systematic errors are canceled, the closure quantities include thermal noise.
The degree of the influence can be recognized by changing the integration time of the data.
\item
Baseline-based systematic errors are not canceled.
This is evident from the definition of closure quantity. Note that most descriptions in textbooks assume that there is no baseline-based systematic error and explain the quantity.
In the case, the closure will not close and 
information of the observed source structure can be lost.
An example is atmospherically-induced incoherence, which is a  non-closing and baseline-based error.
\end{enumerate}
\subsection{Our Hybrid mapping process for the data}\label{Sec:ouh}
Now let us describe the specific analysis we performed.
There are two recording channels in the data, we merged the visibilities from them and made one image result using the hybrid mapping technique, which is commonly used in VLBI imaging and implemented in AIPS (the NRAO Astronomical Image Processing System~\citep{Greisen2003}).
We did not apply any special weighting to the data points, as the EHTC did to mitigate the time variability in the structure of the \SGR based on their GRMHD simulation~(see Section~\ref{Sec:DISC}), but used a very conventional approach.
At every hybrid mapping step, we created several candidate maps by averaging several CLEAN maps, which are composed of CLEAN components, and compared the residuals of closure phases, closure amplitudes, and normalized amplitudes. 
From these, we selected the image that was as consistent as possible with the observed visibility data by checking the difference in closure amounts between the image-converted visibility and the observed one.

We started the hybrid mapping with a point source as the initial image model, following the standard procedure when the source structure is unknown.
The six steps of self-calibration were performed in phase-only mode. However, during each iteration, several resulting images with different imaging parameters (BOX setting, ROBUSTNESS, GAIN, NITER in IMAGR, the task of CLEAN,) were generated and compared to select the next image model. Note that the solution interval ($\mathrm{
SOLINT = 0.15}$) in CALIB (a task of self-calibration)  was set to be sufficiently shorter than the coherence time, following the same approach as in CALIB in~\cite{Miyoshi2022a}.
Data calibration using self-calibration solutions for amplitude was omitted due to the significant intensity variations in \SGR, rendering amplitude adjustments ineffective.
The first and last self-calibration solutions are presented in Appendix~\ref{Sec:snplt}. The first self-calibration solution was 
obtained using the point source model, and the difference between the two channels was smaller than in the case of the \virgoa public data. Additionally, the difference between the last and the first solution was small. We believe that a highly accurate phase calibration was performed in a-priori data calibration step, in comparison to the \virgoa public data.~(Refer to Figure~2 in \cite{Miyoshi2022a} for comparison.)

\begin{figure*}
\includegraphics[width=\textwidth,trim=5 5 45 25,clip]{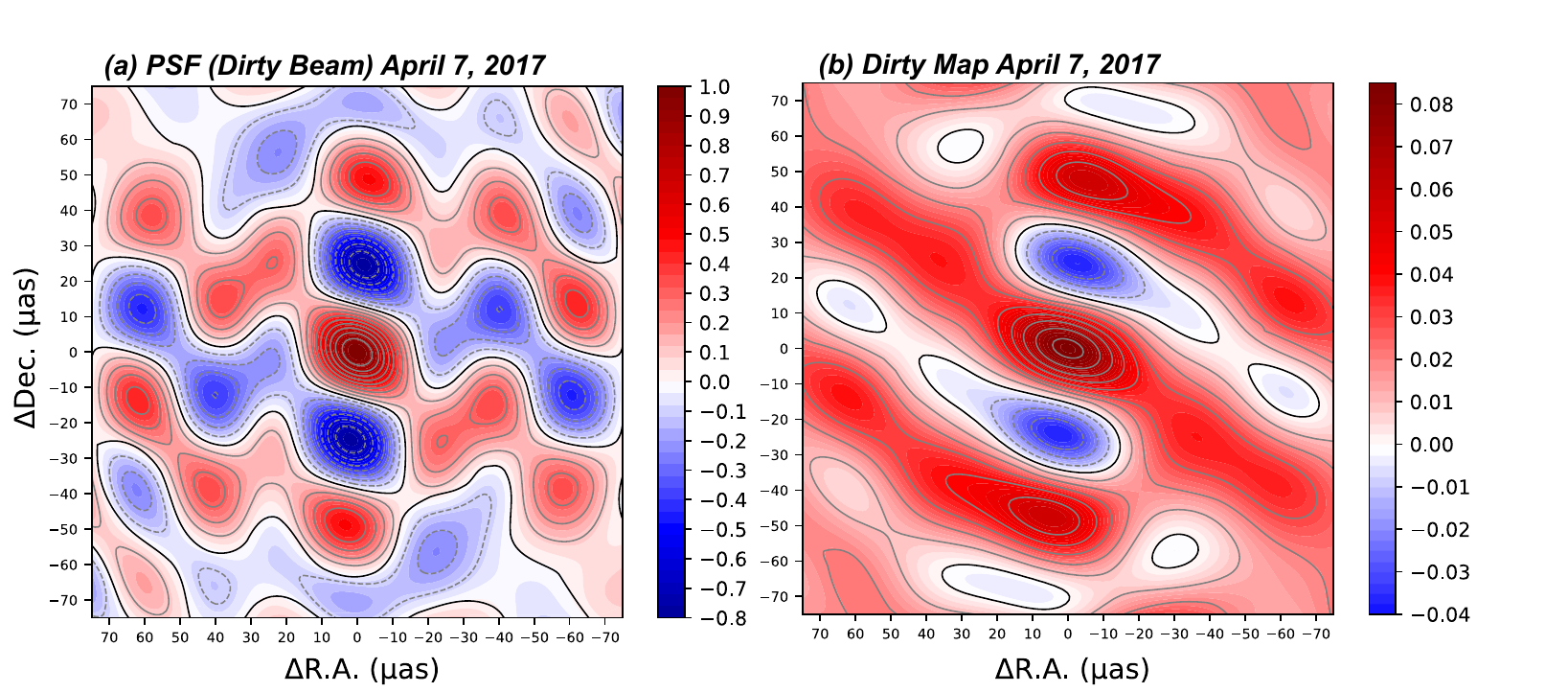}
\caption{
The PSF (dirty beam) and dirty map of the April 7 data. Panel~(a) presents the PSF, with the scale normalized to the height of the main beam. Panel~(b) shows the dirty map obtained after calibration with the solutions of the first self-calibration (phase only) using a point source model. The unit is \textnormal{Jy/beam}.
}\label{Fig:PSF+dritymap47}
\end{figure*}

Figure~\ref{Fig:PSF+dritymap47} shows the PSF (dirty beam) and dirty map of the April 7 data. 
The PSF of the EHT in 2017 is so bumpy that it is beyond comparison with those of common interferometers like Very Large Array (VLA)
VLA and ALMA. The main beam is not sharp and there are many of high level sidelobes and deep dips.
As can be seen from the structure of the PSF, the levels of the sidelobes are of comparable strength to the main beam.
The maximum height of the sidelobes reaches $\mathrm{
49.1\%}$ of that of the main beam, and the deepest point of the negative portion is also large and deep, with a depth of $\mathrm{
-78.1\%}$ of that of the main beam. 
 The spacings between the main beam and the first sidelobes are $\mathrm{\sim 50~\mu \rm as}$. It is challenging to identify the source structure of tens of $\mathrm{\mu \rm as}$ with this PSF, which has a very bumpy structure.

Comparing the dirty map with the PSF, we can see that the central region of the dirty map is much wider and smoother than the main beam of the PSF. This suggests that the central part of the observed source structure is more extended than the spatial resolution ($\mathrm{
\sim 20~\mu \rm as}$). The central structure in the dirty map also exhibits considerable point symmetry with an east-west elongation, indicating possibility that it is a single compact component with one peak rather than one that can be approximated by multiple peaks.
The PSF structures are further discussed in Section~\ref{Sec:uvs} and in Appendix~\ref{Sec:uv46}.

We set the field of view (FOV) for the CLEAN area at \textnormal{
1~\rm mas square}. 
However, in practice, the CLEAN subtraction area is restricted by the BOX setting~($\mathrm{256~\mu \rm as}$ in diameter). 
The size of \SGR~has been previously measured in detail through observations discussed in Section~\ref{Sec:Intro}, and no extended structures, such as a jet, has been detected.
The variation of visibility amplitudes with \UVC distance in Figure 2 of \cite{EHTC2022a} also provides validity that the size of \SGR observed at the EHT 2017 observations is much less than $1~\rm mas$.
We can safely assume that the FOV size for \SGR~is within $1~\rm mas$.

\section{Imaging results}\label{Sec:results}
In this section, we present our final image of \SGR~, and investigate its reliability.
\subsection{Our final image}\label{Sec:foi}
\begin{figure*}
\centering
\includegraphics[width=0.85\textwidth, trim=5mm 2mm 5mm 7mm, clip]{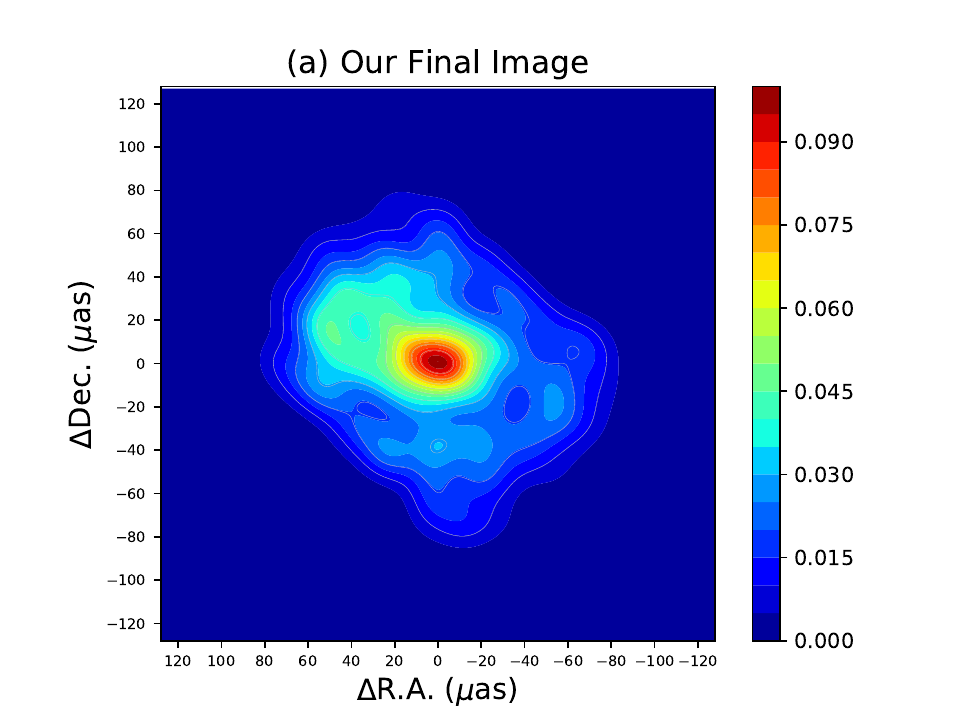}
\includegraphics[width=0.85\textwidth, trim=5mm 2mm 5mm 7mm, clip]{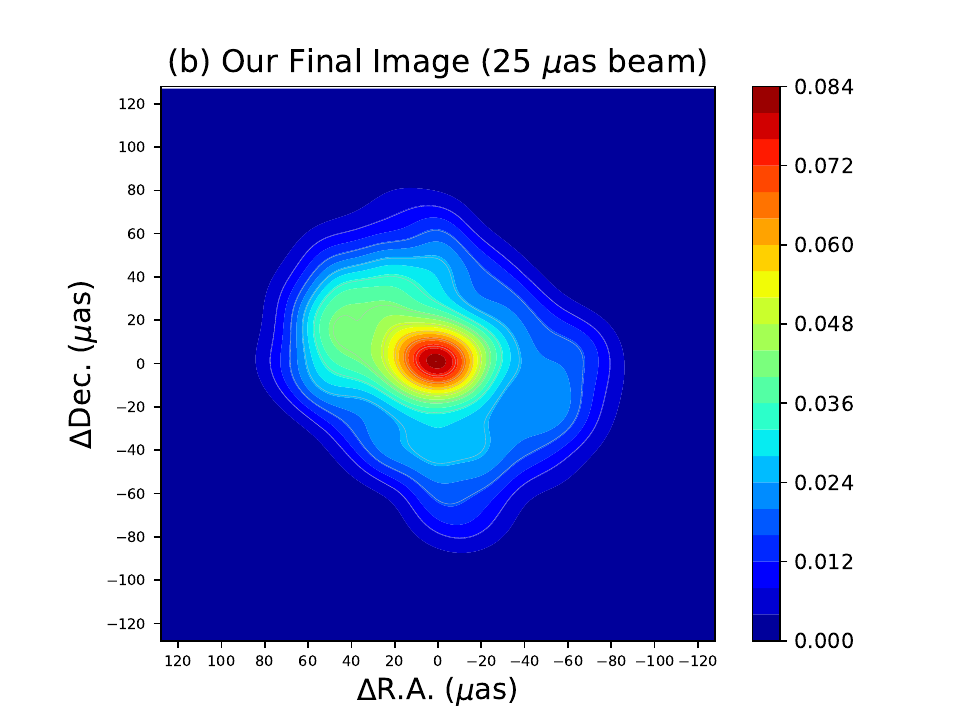}
\caption{
Our final image of \SGR~obtained from our imaging process.
 Panel~(a) 
shows the image convolved with a circular Gaussian restoring beam with a HPBW of $\mathrm{20~\mu\rm as}$, while 
Panel~(b) shows the same image convolved with a circular Gaussian restoring beam with a HPBW of $\mathrm{25~\mu\rm as}$. The brightness unit is $\mathrm{Jy/beam}$. 
}
\label{Fig:ourimages}
\end{figure*}

In contrast to the ring structure reported by EHTC, our final image of \SGR shows an east-west elongation with an asymmetry, the east side appearing brighter than the west side.
We also observe a halo-like extension around the elliptical shape.
This east-west elongation with asymmetry is consistent with previous observations at higher than \textnormal{43~GHz}, as we noted in Section~\ref{Sec:Intro}.

Assuming that the intrinsic angular size of \SGR~is proportional to the observed wavelength~\citep{Johnson:2018}, the shape of \SGR extrapolated to \textnormal{230~GHz} from~\cite{Bower:2014} is an elliptical Gaussian of $\mathrm{66\times 24~\mu \rm as}$ (Half Power Beam Width, HPBW) with a position angle of~$\mathrm{
95~\DEG}$. This expected shape shows a correspondence to that at the \textnormal{50~\%} peak brightness level in our final image, although the PA differs by about~$\mathrm{20~\DEG}$.
Here we make the following working hypotheses: The east-west brightness asymmetry observed in our final image is due to the Doppler boost/de-boost effect caused by the rapid rotation of the accretion disk.

The asymmetric elliptical shape of the accretion disk implies that we are looking at it from roughly an edge-on orientation, with the east side rotating toward us and the west side rotating away from us. 
Assuming that our final image accurately represents the structure of the accretion disk in \SGR, we here attempt to interpret it using the Doppler boosting model.
When observing the edge-on view of the optically-thin accretion disk around a black hole, the Doppler effect results in an asymmetric brightness pattern if the disk is rotating at relativistic velocities.
The brightness ratio $\Delta$ due to Doppler boosting/de-boosting in such a scenario is given by the following equation.

\begin{equation}
\hspace{3cm}\Delta = (\frac{1+\frac{v}{c}\cos\theta}{1-\frac{v}{c}\cos\theta})^3,
\end{equation}

Where {\it c} is the speed of light and {\it v} is the rotation velocity of the disk. Our viewing angle of the disk is $\theta$, where $\mathrm{\theta=0\DEG}$ for the edge-on view and $\mathrm{\theta=90\DEG}$ for the face-on view. 
Figure~\ref{Fig:ourimages}~(a)
, we can observe the maximum bright spot on the east side ($\mathrm{2.1701\times 10^{-4}Jy/Beam}$, at ($\mathrm{0,-2~\mu \rm as}$)) and the dark spot on the west side ($\mathrm{3.8635\times 10^{-5}Jy/Beam}$, at ($\mathrm{-37, -17~\mu \rm as}$)). We assume that the brightest spot is Doppler-boosted, while the darkest spot is Doppler-de-boosted.

If the relative ratio is $\mathrm{\Delta=\frac{2.1701\times 10^{-4}}{3.8635\times 10^{-5}}=5.617}$, then
\begin{equation}
\hspace{3cm}\frac{v}{c}\cos\theta =  0.438.
\end{equation}
Assuming a distance to the Galactic Center of $\mathrm{D_{GC} = 8 kpc}$ and a black hole mass of $\mathrm{4\times10^6M_{\odot}}$ in \SGR~, 
the angular distance of $\mathrm{21.3~\mu \rm as}$, half the separation between the boost and de-boost spots, corresponds to $\mathrm{\sim 2.11 }$\textnormal{\RS}~in physical length. 
This length is already smaller than \textnormal{3~\RS}, the diameter of the last stable circular orbit of a Schwarzschild black hole. 
It probably means that the \SGR~black hole is a rotating Kerr black hole.
At this radius from the black hole, the general relativistic circular velocity is $\mathrm{v_{GR} \sim 0.607~\rm c}$, and the viewing angle is estimated to be $\mathrm{\theta \sim 43.8 \DEG}$. The observational data likely contains information on the accretion disk emission in the range of $\mathrm{\sim 2}$ to \textnormal{a few \RS}~ from the black hole.\\
~There is a variety of observational estimations of the viewing angle of the accretion disk of \SGR~, ranging from edge-on to face-on. 
Recent GRAVITY observations report an angle of $\mathrm{
i = 160\pm 10\DEG}$\citep{GRAVITY:18}, which is close to face-on, 
while radio observations are often close to edge-on. \cite{Cho:22} suggest that the viewing angle is less than or equal to $\mathrm{
30\sim40 \DEG}$, 
while higher inclinations of around $\mathrm{50~\sim68\DEG}$ have been proposed by \cite{Dexter:2010},~\cite{Broderick:2011}, and \cite{Wang:2013}. 
\cite{Miyoshi:11} have shown that the angle is nearly edge-on, with a range of $\mathrm{87\sim~87.7\DEG}$ from peak motion in the SMI analysis.\\
~The above discussion of the inner accretion disk of \SGR~was conducted assuming that our final image is correct. 
Again, this is based on the assumption that the fine structure seen in Figure~\ref{Fig:ourimages} ~(a) is realistic. Figure~\ref{Fig:ourimages} ~(b) shows another image where a $\mathrm{
25~\mu \rm as}$ Gaussian beam is used to convolve the CLEAN components. 
Note that in this image, the dark point on the west side, which is assumed to be the Doppler de-boost point, disappears, 
so the reliability of the fine structure seen in 
Figure~\ref{Fig:ourimages}~(a) is not very high.

\subsection{Reliability of our final image}\label{Sec:Reliability}

To assess the reliability of our final image, we computed the residuals of normalized amplitudes, closure phases, and closure amplitudes, 
and compared them to those of the EHTC image.
As for the EHTC image, we digitized Figure 3 in \cite{EHTC2022a} to obtain each coordinate value and the intensity at that coordinate.
After modifying the intensity at each point, assuming an overall intensity of \textnormal{1~Jy}, this numerical image data was input into the AIPS task, UVMOD, to create a visibility set with \UVC coverage of the EHT 2017
\footnote{The EHTC imaging pipelines of \SGR~have not been made public, unlike the case of \virgoa.}.

There is no clear difference between our final image and the EHTC ring image when considering the residuals about the closure phase and the closure amplitude.
In terms of the residuals of the normalized amplitudes, our final image exhibits half the residuals in the case of the EHTC ring image. However, it is worth noting that both our final image and the EHTC ring image have significantly larger residuals than those of \virgoa~\citep{Miyoshi2022a}. Presumably this is an effect of the rapid  intra-day variability of \SGR.

\subsubsection{Residuals of normalized amplitude of our final image compared to those of the EHTC ring}\label{Sec:NAR}
\begin{figure}
\begin{center} 
\includegraphics[width=\columnwidth,trim=35 3 34 20, clip]{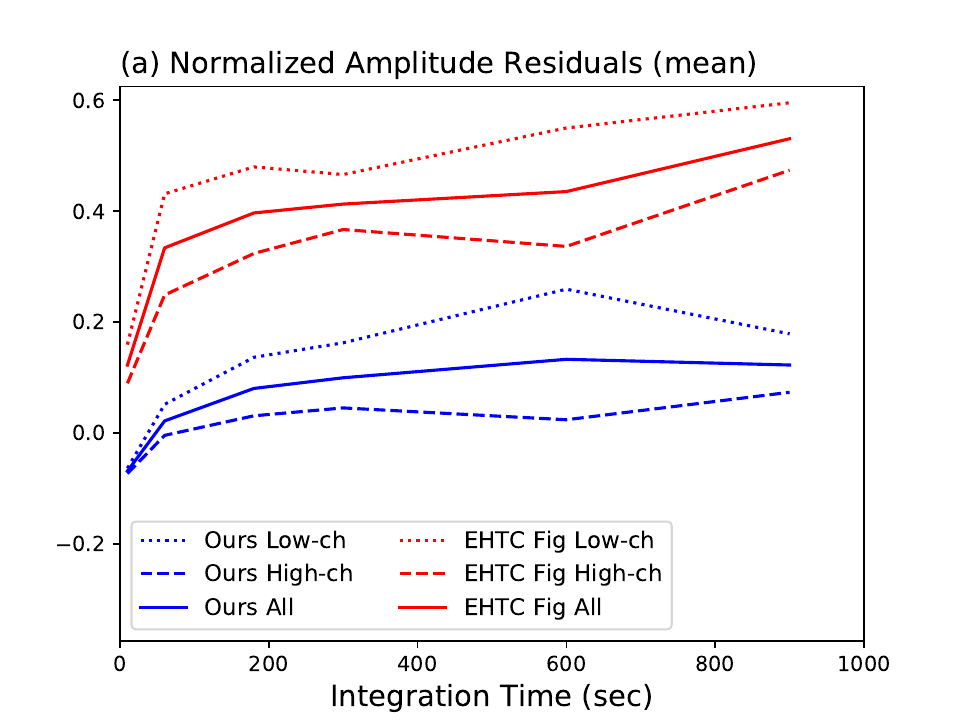}
\includegraphics[width=\columnwidth,trim=35 1 34 0, clip]{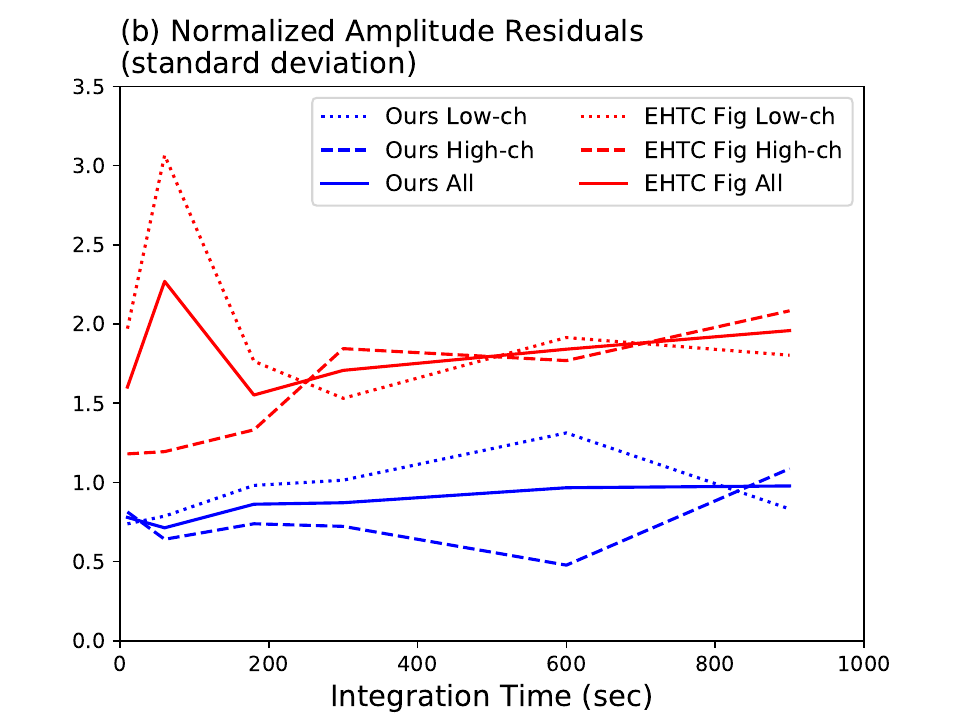}
\end{center} 
\vspace{-0.5cm}
\caption{
Statistics of the residuals of the normalized amplitudes. 
Panel~(a) shows the means of the residuals of the normalized amplitudes, while Panel~(b) shows their standard deviations. 
The red lines represent the statistics (mean, standard deviation) of those for the EHTC ring image, and the blue lines represent those for our final image. 
The solid line shows the statistics of the residuals for all data, and the dashed line shows the statistics for the low and high recording channels, respectively.
}
\label{fig:NRamp-resi}
\end{figure}

Figure~\ref{fig:NRamp-resi} shows the residuals of normalized amplitudes
(Section~\ref{vischk}, Appendix~\ref{Sec:NAR-def})
; EHTC imaging teams also use this measure in their analysis. 
Our final image consistently exhibits smaller means and standard deviations in the residuals than the case of the EHTC ring image, regardless of the integration time between 
$\mathrm{T_{\text{int}} = 10~sec}$ and 
$\mathrm{T_{\text{int}} = 900~sec}$.
The mean of the residuals of our final image range from 0 to 0.2, whereas those of the EHTC ring image range from 0.2 to 0.6. 
The standard deviations of our final image are around 1, while those of the EHTC ring image range from over 1 to 3.
For example, our final image shows the residual of normalized amplitudes 
$\mathrm{NR_{ours} = 0.080 \pm 0.397}$, 
while the EHTC ring shows the residual of normalized amplitudes 
$\mathrm{NR_{EHTC} = 0.862 \pm 1.552}$, for $\mathrm{T_{int} = 180~sec}$.
However, compared to the residuals in the case of~\virgoa
~(Section~\ref{Sec:caseM87}), the values of the residuals of the normalized amplitudes obtained for both \SGR images are remarkably large.
At least, we are confident that these large residuals are not due to miscalculations on our part, as \cite{EHTC2022d} shows residuals from their analysis that are consistent with our own residual analysis results for the EHTC ring image.
The larger residuals than those of \virgoa are certainly due to short-term variations in the intensity and structure of \SGR. This is a result of the forced attempt to represent a variable object in a single static image.

A characteristic feature is the behavior of the residuals with respect to the integration time.
In our final image, the values do not vary significantly with integration time, while for the EHTC ring image, both the mean and standard deviation seem to increase gradually with integration time.

Typically, residuals caused by thermal noise decrease with integration time. Since this is not observed in this data set, it can be concluded that the residuals are not due to thermal noise but rather some systematic errors. The residuals about closure phase and closure amplitude described below exhibit similar behavior, suggesting that something other than thermal noise is predominant.

A possible reason for the larger residuals in the case of the EHT ring image with increasing integration time could be insufficient coherence recovery. 
If the image is an inadequate model, the calibration of the data by the self-calibration solution will be poor, resulting in insufficient removal of the effect of phase variations due to atmospheric vapor, leading to smaller amplitude values indicated by the data and potentially larger residuals in the normalized amplitude.
The observed elevation angles of \SGR at each station are not as high as might be expected. Those of ALMA and the Atacama Pathfinder Experiment~(APEX) at the Chili site reach a maximum elevation angle of about $\mathrm{80 \DEG}$, while LMT, the James Clerk Maxwell Telescope~(JCMT)
, and the Submillimeter Array ~(SMA) have a maximum elevation angle of about $\mathrm{40 \DEG}$.
The Submillimeter Telescope Observatory in Arizona~(SMT, AZ) and 
the Pico Veleta~(PV) stations have an elevation angle of less than $\mathrm{30 \DEG}$.
The SPT station is located at the South Pole, so the observed elevation angle is almost constant at $\mathrm{30 \DEG}$.
The lower the elevation angle, the larger the effect of atmospheric phase variations.

%
\subsubsection{Residuals of closure phase of our final image compared to those of the EHTC ring}\label{Sec:RCP}
The closure phase, 
as shown in Section~\ref{visclo} and Appendix~\ref{Sec:RCP-def}, 
is a quantity that is immune to systematic errors resulting from antennas and depends solely on the brightness distribution of the source
as long as baseline-based errors do not exist.

The discrepancy between the observed visibility and the visibility derived from the reconstructed image serves as a crucial metric to evaluate the consistency between the reconstructed image and the observed source structure.

\begin{figure}
\begin{center} 
 \includegraphics[width=\columnwidth,trim=35 3 34 25, clip]{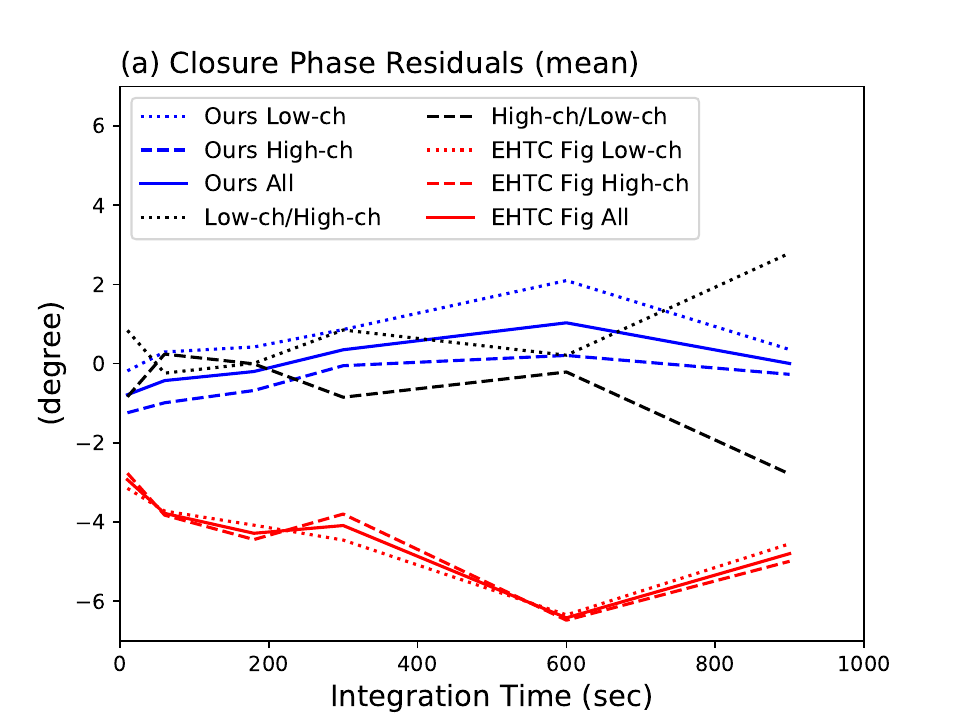}
 \includegraphics[width=\columnwidth,trim=35 1 34 15, clip]{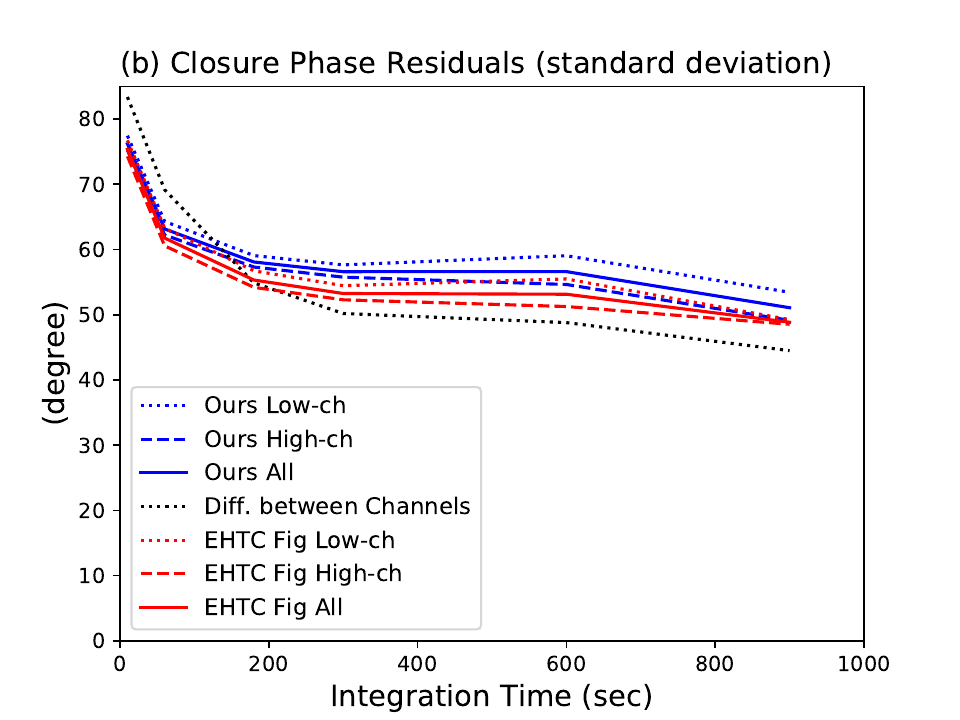}
\end{center} 
\vspace{-0.5cm}
\caption{
Statistics of the residuals of the closure phases.
 Panel~(a) shows the means of residuals of the closure phases.
 Panel~(b) shows the standard deviations of them.
The red lines show the statistics (mean, standard deviation) of the residuals of the closure phases for the EHTC ring image, and the blue lines show them for our final image.
The statistics of the residuals for all data are shown by the solid lines.
There are two recording channels (low and high).
The respective statistics of the residuals are shown by the dotted lines.
In addition, the black dotted lines show the difference between the two recording channels.
}
\label{fig:res-closureP-sta}
\end{figure}

Here, we compare the residuals of the closure phases about our final image and the EHTC ring image~(Figure~\ref{fig:res-closureP-sta}). 
To give an example, in the case of 
$\mathrm{T_{int} = 180~sec}$, for our final image, the residuals are 
$\mathrm{Res_{cp} = 0.2 \pm 58.1^\circ}$, 
on the other hand, the residuals for the EHTC ring are $\mathrm{Res_{cp} = -4.3 \pm 55.3^\circ}$.

The means of the residuals of our final image are close to zero regardless of the integration time.
On the other hand, although we have to admit that the values are not far from zero, considering the size of the standard deviations, those of the EHTC ring image are always a few degrees away from zero, regardless of the integration time. 
 On this basis we believe that our final image is probably a bit closer to the actual source structure.

In both cases, however, the standard deviations do not decrease with the integration time. If the error is due to thermal noise, the values should decrease in proportion to the power of -0.5 of the integration time ($\mathrm{\propto~T_{intg}^{-0.5}}$).

We speculated that the reason why the mean is close to zero and the standard deviations of $\mathrm{\sim50~\DEG}$ are constant regardless of the integration time may also be due to the time variation of the \SGR source structure.
To confirm this picture, we investigated the difference in closure phase between the two recording channels, which should theoretically show the same value as they are recorded simultaneously, despite the \SGR~'s strong short-term variation.
Thus, the difference in closure phase between the channels should be zero on average. Since only thermal noise is present, the standard deviation should decrease in proportion to the power of -0.5 of the integration time.

The following detailed calculations show that a small difference in the observed frequency setting between the channels can be negligible.
The observed frequencies of the two channels are only slightly different, with a frequency difference of \textnormal{0.88~\%} (\textnormal{227.1} and \textnormal{229.1~GHz}). 
Certainly there is a frequency dependence of the structure of the \SGR~, but this difference by frequency is less than a mere \textnormal{1~\%}.
Applying \cite{Johnson:2018}'s estimate here, the difference in angular broadening due to intervening plasma is 
$\mathrm{0.4 \times~0.2~\mu \rm as}$,
For the intrinsic size, the difference is only 
$\mathrm{0.46~\mu \rm as}$ in size. 
These are much smaller than the spatial resolution of the observations ($\mathrm{\sim20~\mu \rm as}$), and thus the visibility differences between the channels due to the source 
structure are essentially negligible.

However, unexpected results were obtained, suggesting the existence of internal inconsistencies in the data information.
The standard deviation of the closure phase difference between channels is larger than expected.
For example, the difference for 
$\mathrm{T_{\text{int}} = 180~sec}$ integration is 
$\mathrm{0.007~\pm~54.9\DEG}$, which is not much different from the value for the closure phase residual shown above. 
Surely, the average shows nearly zero as expectation while the standard deviation does not decrease much even when the integration time is changed.
If the source structure differs significantly between the frequency channels, the average of the difference between the two channels will not be zero. However, since the observed average is around zero, it cannot be interpreted that the observed source structure has a significant frequency dependence.
It presumably mean that the data have an internal inconsistency with respect to the closure phase.

Figure~\ref{fig:res-closureP-sta} also shows the closure phase difference between channels.
Inconsistencies in the closure quantities suggest a possibility that the data do not contain correct information about the source structure.
To explore this issue in depth, we examined all the \SGR~observational data publicly available from EHTC. The differences between the two recording channels of these data are described in the Appendix~\ref{Sec:betweeIFs}.

%
\subsubsection{Residuals of closure amplitude of our final image compared to those of the EHTC ring}\label{Sec:RCA}
We here examine the residuals of the closure amplitudes. 
The closure amplitude, as shown in Section~\ref{visclo} and Appendix~\ref{Sec:RCA-def}, 
is a quantity that is insensitive to systematic errors from each antenna.
It also serves as a metric for evaluating the consistency between the reconstructed image and the actual structure of the observed source.

\begin{figure}
\begin{center} 
\includegraphics[width=\columnwidth,trim=35 3 34 25, clip]{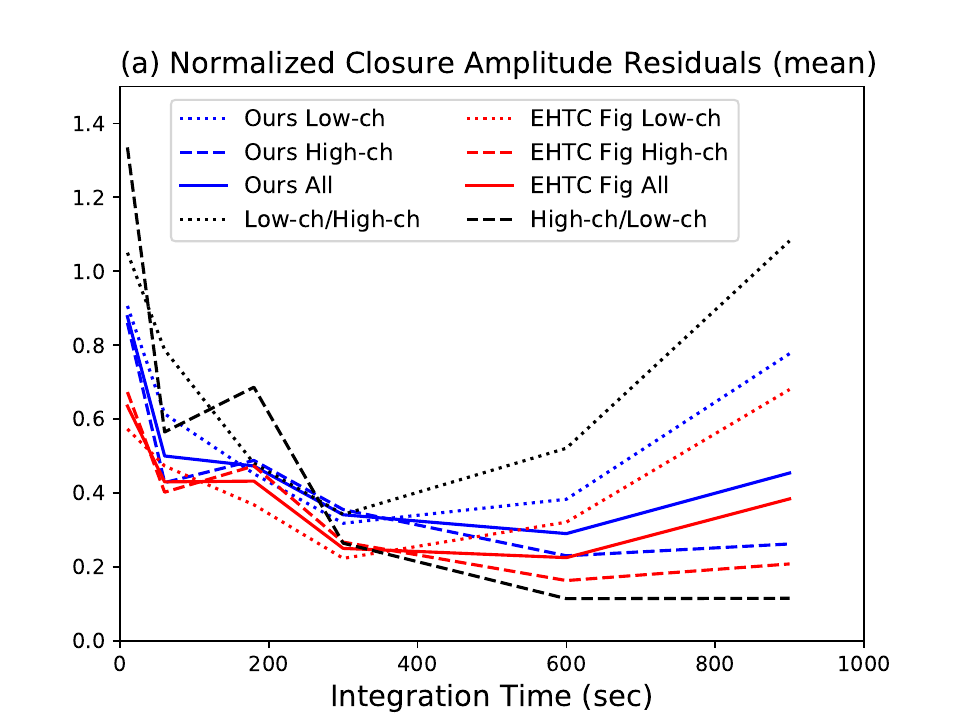}
\includegraphics[width=\columnwidth,trim=35 3 34 5, clip]{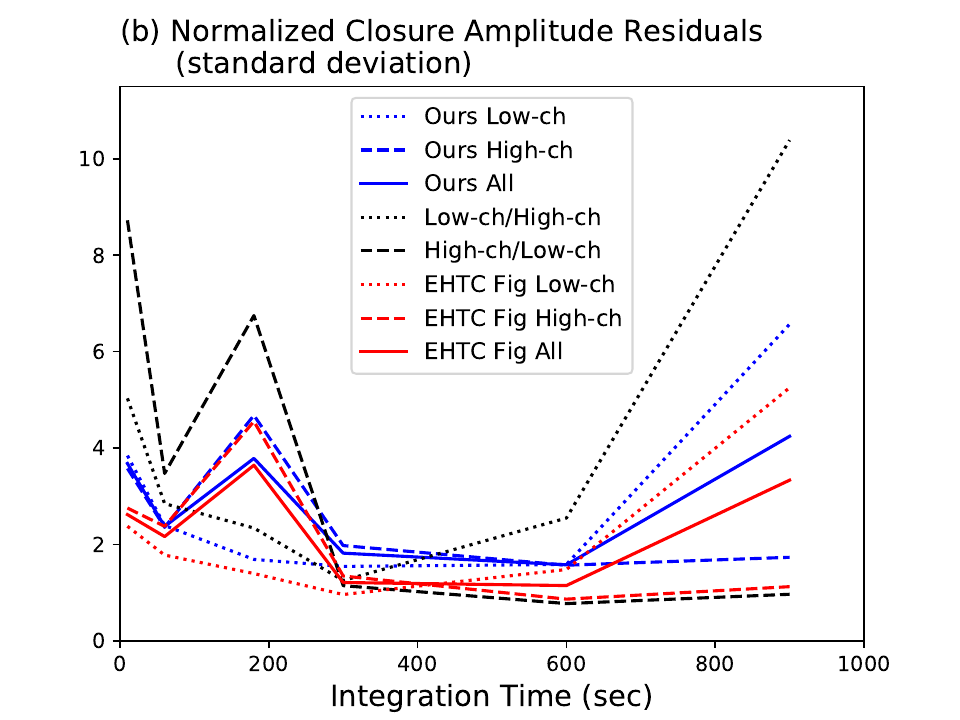}
\end{center} 
\vspace{-0.5cm}
\caption{
Statistics of the residuals of the normalized closure amplitudes.
Panel~(a) shows the mean values of the means, while 
Panel~(b) shows the corresponding standard deviations. 
The solid lines represent the statistics (mean and standard deviation) of the residuals of the normalized closure amplitudes for all data. 
The red lines indicate the statistics for the EHTC ring image, while the blue lines correspond to our final image. 
The dotted lines represent the statistics of the residuals for the low and high recording channels separately. 
The black dotted lines represent the difference between the two recording channels.
}
\label{fig:res-closureA-sta}
\end{figure}

In Figure~\ref{fig:res-closureA-sta}, we present the actual residuals of the normalized closure amplitudes.
For example, our final image shows the residual of 
$\mathrm{Res_{ca} = 0.473 \pm 3.784}$, while the EHTC ring shows the residual of $\mathrm{Res_{ca} = 0.432 \pm 3.643}$, for 
$\mathrm{T_{int} = 180~sec}$. The difference in closure amplitude between channels is shown for reference: 
For ${\mathrm{T_int} = 180~sec}$, 
    $\mathrm{Res_{ca} = 0.480 \pm 2.339}$ based on the high channel, $\mathrm{Res_{ca} = 0.686 \pm 6.746}$ based on the low channel.

In both of our final image and the EHTC ring image, the residuals of the normalized closure amplitudes are not negligible. 
Again, we observe that the values do not necessarily decrease with integration time. 
For integration times up to \textnormal{300 sec}, both the mean and standard deviation generally decrease. 
However, for integration times above \textnormal{300 sec}, both increase. Moreover, it is noteworthy that the maximum values are sometimes obtained at the integration time of \textnormal{180 sec}.

In Figure~\ref{fig:res-closureA-sta}, not only the residuals of the normalized closure amplitudes, also we show the differences of those between the two recording channels. 

The details are described in Appendix~\ref{Sec:CA2}
If the differences in the closure amplitudes between the channels found here mean that the closure amplitudes contain some kind of systematic errors, it is difficult to select the optimal image based on a smaller residual of amounts.

\hspace{1cm}
\subsubsection{PSF effect on our final image}\label{Sec:ourPSF}
\begin{figure*}
\centering
\includegraphics[width=\textwidth]{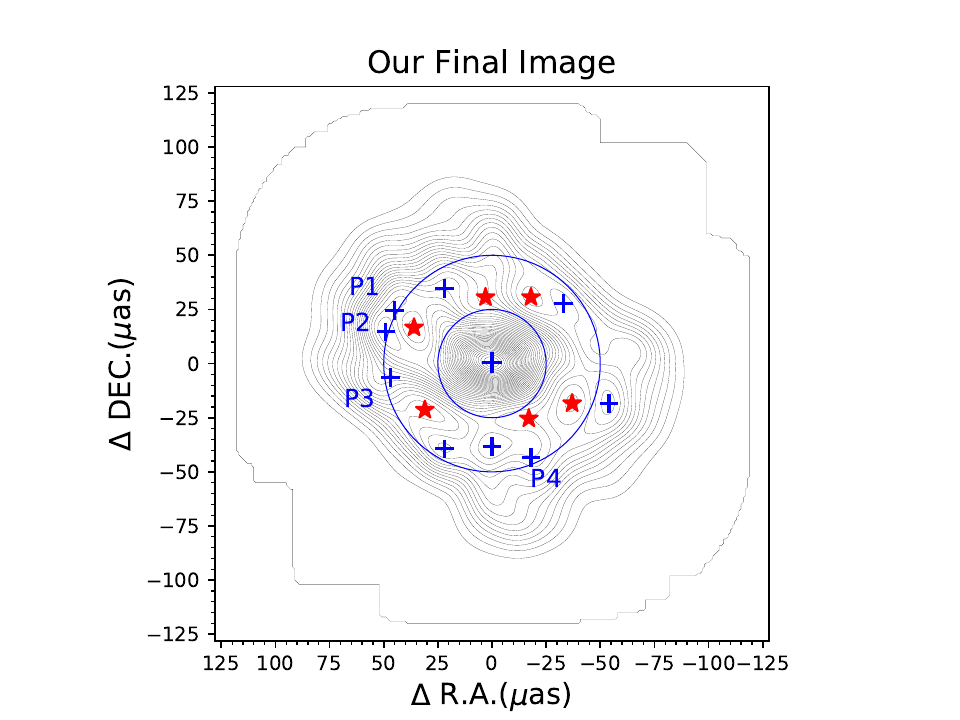}
\caption{
PSF effect on our final image.
Local maxima are indicated by blue crosses and local minima are indicated by red asterisks. 
P1, P2, P3, and P4 are local maxima located $\mathrm{
50~\pm 2.5 \mu \rm as}$ from the central peak.
Contour lines represent each \textnormal{2~\%} interval from the peak to the zero level in 50 increments. 
Circles with diameters of \textnormal{25} and $\mathrm{50~\mu \rm as}$ are drawn centered on the central peak position.
}
\label{Fig:PSFourI}
\end{figure*}

 In general, it is very difficult to obtain an image completely free from PSF structure. 
We examined our final image for features that might be due to the PSF structure.
The structure of $\mathrm{\sim 50~\mu \rm as}$ spacing is prominent in the PSF of the EHTC array in 2017 for \SGR. Most prominent is the spacing between the main beam and the first sidelobe, which is $\mathrm{
49.09~\mu \rm as}$.
Figure~\ref{Fig:PSFourI} shows our final image with contour expression.
Several local maxima can be seen at a distance of about $\mathrm{
50~\mu \rm as}$ from the central peak. In particular, P1, P2, P3, and P4 are located at a distance of 
$\mathrm{50~\pm~2.5~\mu \rm as}$ from the peak. This is probably due to the effect of the PSF structure. However, they do not have a significant effect on the overall brightness distribution of our final image.
\subsubsection{Summary of the reliability of our final image}\label{Sec:SRI}
To summarize the reliability of our final image in terms of residuals in closure phase and amplitude, we found no significant difference between our final image and the EHTC ring. However, the residuals of the normalized amplitudes in our final image were about two times better, indicating that our final image is more self-consistent in data analysis.

In this investigation, we found a probably serious problem with the data quality itself, which is completely independent of the time variability of \SGR. There may be some internal inconsistency in the EHT data processing, including the correlation processing. 
If the closure quantity is not preserved, there is a possibility that the image information is corrupted, although the degree is unknown.
Since both our final image and the EHTC ring image show the same degree of closure residuals, it is possible that the data are corrupted to the degree that the shape, which is about $\mathrm{50~\mu \rm as}$ in size,  cannot be determined from the residual analysis of the closure quantity.
~As shown in Appendix~\ref{Sec:betweeIFs}, an unexplained closure discrepancy between recording channels is present in all EHTC data. 
Identifying the cause of these unknown errors using only the EHT public data is 
 extremely challenging. 
The EHT public data have reduced time resolution and loss of bandwidth and polarization characteristics due to the averaging and combining of multiple channels. 
 Checking the raw output data of the correlator is necessary to identify the cause of discrepancies in the closure quantity. If the recorded data are available, the correlation process should be reproduced.
%
\section{Relation between the EHTC ring shape and the visibility data}\label{Sec:the EHTC-ring}
Here we explain that for the EHT 2017 observations of \SGR~, a ring with a diameter of about $50~\mu \rm as$ can arise due to the PSF structure.  
This is mainly because the PSF structure has its highest sidelobes at $\mathrm{50~\mu \rm as}$ from the main beam, with their midpoints being the deepest negative minima.
\subsection{The \UVC coverage and the corresponding PSF of the EHT array in 2017 for \SGR}\label{Sec:uvs}
\begin{figure}
\begin{center}
\includegraphics[width=\columnwidth, trim=8 3 40 25, clip]{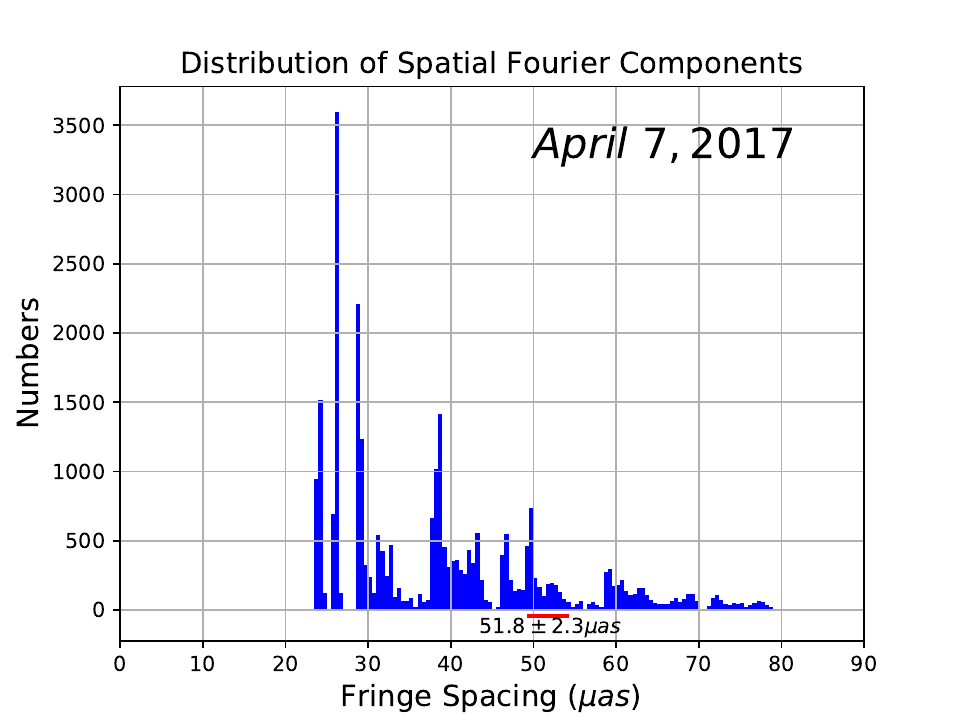}
\end{center}
\caption{
Distribution of the sampled data from all baselines (April 7, 2017). The horizontal axis shows the fringe spacings of the sampled visibility data (spatial Fourier components) in $\mathrm{\mu\rm as} $ units. The vertical axis shows the number of samples. 
The red line segment shows the diameter of the ring as measured by the EHTC ($\mathrm{d = 51.8 \pm 2.3~\mu\rm as}$).
In addition to these, there are spatial Fourier components from about \textnormal{2.2 arcsec} to \textnormal{3.2 arcsec} in the sampled data. Here, such large samples are omitted from this plot. 
}\label{Fig:uvs47}
\end{figure}

Here, we examine the \UVC sampling of the EHT array in the 2017 \SGR~observations.
The EHTC paper~\citep{EHTC2022a} shows the \UVC coverage in the \UVC plane. 
Unfortunately, this type of figure, which is usually shown in many papers, does not tell us about the amount of data. So in Figure~\ref{Fig:uvs47} we show the distribution of data sampling versus fringe spacing.
There are samplings from the minimum fringe spacing of $\mathrm{24~\mu\rm as}$ to $\mathrm{80~\mu\rm as}$ in the plot, but it can be seen that the number of samples varies considerably with fringe spacing.
Of concern is the presence of two complete unsampled voids below $\mathrm{30~\mu\rm as}$.

Based on the \UVC distribution of the \SGR~observations, we anticipated 
the PSF to have a pronounced bumpy structure with less than $\mathrm{30~\mu\rm as}$ intervals.
Since the EHTC ring image size of \SGR~is measured to be $\mathrm{50~\mu\rm as}$ in diameter~\footnote{The diameter of the black hole shadow measured by the EHTC is $\mathrm{d = 48.7\pm 7.0~\mu \rm as}$ and the diameter of the bright, thick ring to be $\mathrm{d= 51.8\pm 2.3~\mu \rm as}$ \citep{EHTC2022a}}
, and not around $\mathrm{30~\mu\rm as}$, we hoped this time that the EHTC ring of \SGR~is not due to the PSF structure, but to the actual feature of the source.

\begin{figure}
\begin{center} 
\includegraphics[width=\columnwidth]{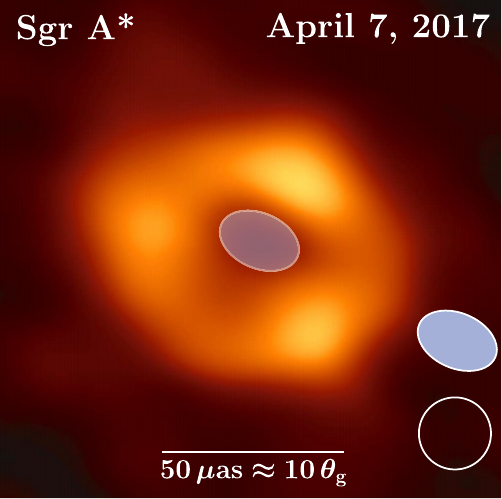}
\end{center} 
\caption{Comparison of the default restoring beam and the EHTC image.
The default restoring beam is an ellipse with $\mathrm{FWHM=23.0\times15.3~\mu \rm as}$ and $\mathrm{PA= 66.6\DEG}$, which shown as a blue ellipse in the panel.
The white circle shows the restoring beam used by EHTC to make their images.
The original image is taken from Figure~3 in~\citet{EHTC2022a}}
\label{fig:default-beam}
\end{figure}

\begin{figure*}
\centering
\includegraphics[width=1.05\textwidth,trim=10 5 15 25,clip]{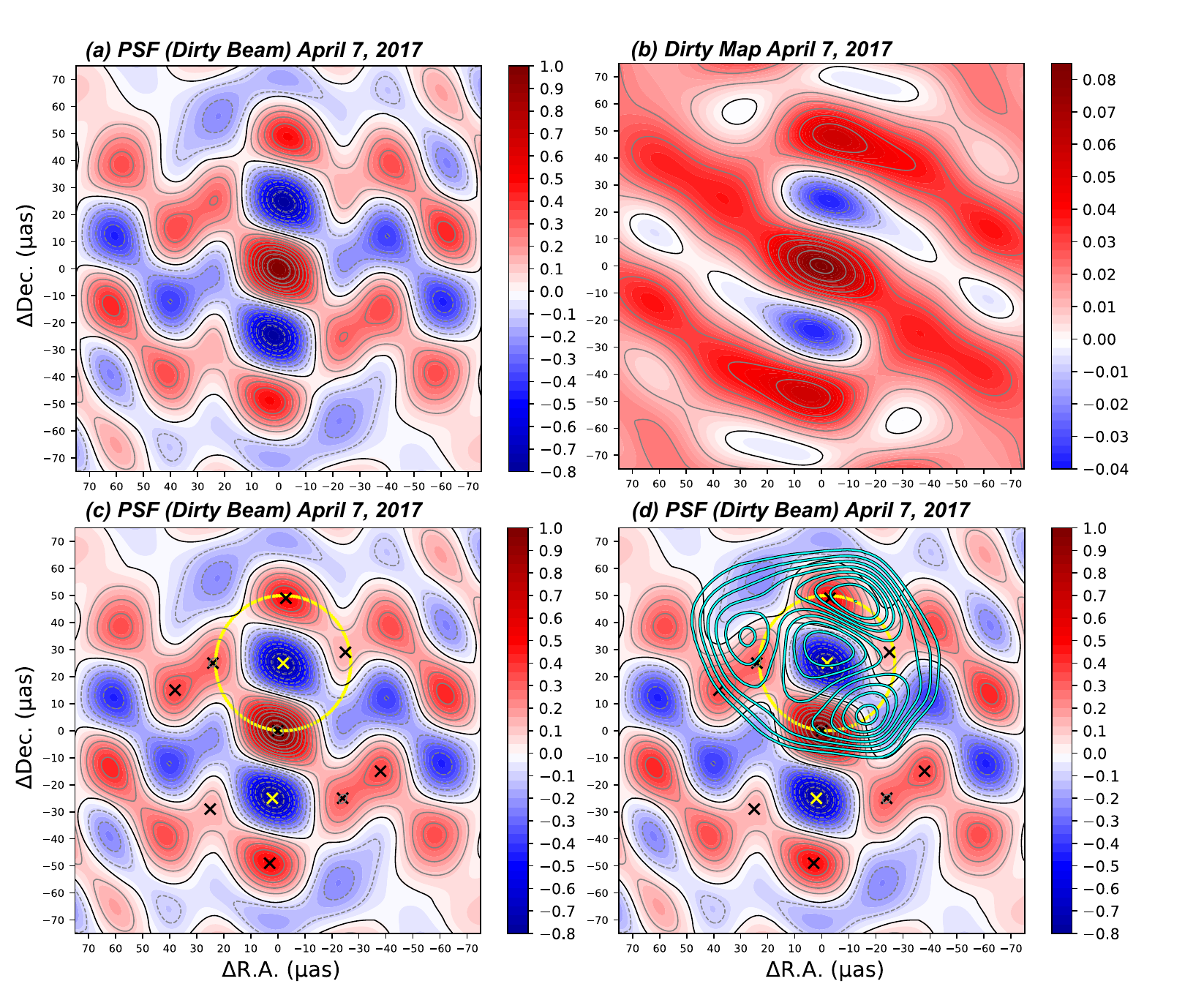}
\caption{
The point spread function (dirty beam) and the dirty map of the EHT array (2017) on the second day of \SGR~observations. 
Panels~(a), (c), and (d) show the point spread function (dirty beam) of the observations conducted on April 7, 2017. 
Panel~(b) presents the dirty map obtained by Fourier transforming the visibility data. 
In Panel~(c), black x-marks represent the peak positions near the center, while yellow x-marks depict the deepest minimum positions. 
The yellow dotted line indicates the circle with a diameter of $\mathrm{50~\mu \rm as}$ centered at the deepest minimum (north) in the dirty map. 
In Panel~(d), we overlay the EHTC ring image (blue contour lines) on the dirty beam. 
The contour intervals are set at every \textnormal{10~\%} of the peak value.
The width of the EHTC ring is wider than that of each peak's structure in the PSF. 
This suggests that the observed source is not a point source, but has a significant size, and thus is broadened. 
The intensity of each PSF peak (or minimum) is defined as the ratio of the main beam height to that of each peak, while the dirty map's intensity is in \textnormal{Jy/BEAM} (in the case that the beam size is $\mathrm{20~\mu \rm as}$).
}
\label{Fig:psf47}
\end{figure*}

In the actual PSF structure, noticeable up-and-down structures with a scale of about $\mathrm{50~\mu \rm as}$ are observed (Figure~\ref{Fig:psf47}). Therefore, examining the \UVC coverage alone is insufficient to fully understand the PSF structure. It is essential to illustrate the PSF structure itself, rather than just the \UVC coverage plot, in scientific papers.
The detailed characteristics of this PSF are outlined below. 
The main beam is formed at the center of the PSF structure, and a Gaussian fit of the main beam shape yields a default restoring beam
of $\mathrm{(FWHM) =23.0\times15.3~\mu \rm as}$ and 
$\mathrm{PA= 66.6\DEG}$~\citep{EHTC2022c}.
Based on this size measurement, the spatial resolution of the imaging are defined to be $\mathrm{20~\mu \rm as}$ by EHTC. This is a quite natural measurement, and is also confirmed to be appropriate in our analysis.
For reference, a comparison of the default restoring beam and the one used by the EHTC and their relationship to the image obtained by the EHTC are shown in Figure~\ref{fig:default-beam}. The shape and size of the default restoring beam just fits into shadow, the central hole of the EHTC ring.

When checking the PSF structure, it is important to consider not only the shape and size of the main beam but also other structures that appear. The first thing to consider is the height of the maximum sidelobes and their positions relative to the main beam. In Figure~\ref{Fig:psf47}, the highest sidelobes are located at ($\mathrm{-3, +49~\mu \rm as}$) and 
($\mathrm{+3, -49~\mu \rm as}$) relative to the main beam. 
Both of these two sidelobes are $\mathrm{49.09~\mu \rm as}$ away from the main beam.
The distances are equal to the diameter of the EHTC ring.
The peak heights of the sidelobes reach $\mathrm{+49.1~\%}$ of that of the main beam, which is orders of magnitude higher than those found in the PSFs of purpose-designed arrays.
If the amplitude calibration of the data is insufficient, false peaks higher than the true source peak can appear at distances equal to that between the main beam and the sidelobes.

Another important point to note is the presence of deep negative values in the PSF. 
This is also because there are insufficient spatial Fourier components  contributing to the PSF structure.
For the data set sampled by the EHT array in 2017, the PSF structure exhibits a very deep negative region.
The blue area in Figure~\ref{Fig:psf47} shows where the level is negative. 
The deepest minima in the PSF structure appear at 
($\mathrm{-2, +25~\mu \rm as}$) and 
($\mathrm{+2, -25~\mu \rm as}$), with a distance of 
$\mathrm{25.08~\mu \rm as}$ from the main beam. 
This distance is the same as the radius of the bright thick ring measured by the EHTC ($\mathrm{r=\frac{51.8\pm2.3}{2}=25.9\pm1.15~\mu \rm as}$).
It is also notable that the depth of the deepest minima reaches $\mathrm{-78.1~\%}$ of the height of the main beam. PSF structures such as that of ALMA do not have such deep dips.

In summary, the PSF of the EHT array in 2017 for \SGR~does not form a sharp and high main beam, but rather forms a bumpy structure with high sidelobe peaks and several negative deeps over a wide area. Additionally, the up-and-down structure in the PSF has a typical spacing of about $\mathrm{50~\mu \rm as}$. 
As shown in Figure~\ref{Fig:psf47}, if we draw a circle with a diameter of $\mathrm{50~\mu \rm as}$ centered on the lowest point between the main beam and a highest sidelobe (located in the north), a group of peaks will be located along the circumference of the circle.
 In our view, therefore, there is a high possibility of creating artifacts of $\mathrm{50~\mu \rm as}$ size in the imaging results
from the \SGR~data obtained from the EHT array in 2017.
%
%
\subsection{Rings from the simulated data of other structures}\label{Sec:simData} 

Here we systematically investigate the possibility that, due to the PSF structure of EHT 2017 array for \SGR~a ring-like structure can be created as an artifact.
We created two simulated data sets by using a AIPS task, UVMOD. One corresponds to that with no source structure, but with a realistic amount of noise. The other is that with a point source structure without noise. Both have the same \UVC coverage as for the \SGR observations of the EHT array (2017).

Actual observational data usually contain information about the structure of a finite size object with noise errors.
The simulated data used here are the farthest extremes from such actual observational data, and are obviously completely different from the ring structure data.
The "point without noise" data contain information about a very simple structure, and since there is no noise, it can be immediately inferred that they are point source data (visibility amplitude is constant independent of the projected baseline length, closure phase is always zero, and closure amplitude is always 1).
In the case of "no source structure with noise" data, the visibility phase is randomly and uniformly distributed over $\mathrm{0}$ to $\mathrm{2~\pi}$, so we can conclude that the data is dominated by noise and it makes no sense to start the imaging process.
In the actual data analysis, the characteristics of the visibility data are examined before the imaging process, and it can be understood that the simulated data here do not show any rings without the imaging process.
What we wish to show here is the strong power of self-calibration and FOV restriction in imaging on the sparse \UVC EHT 2017 data.

\begin{figure*}
\includegraphics[width=\textwidth,trim=55 10 50 5,clip]{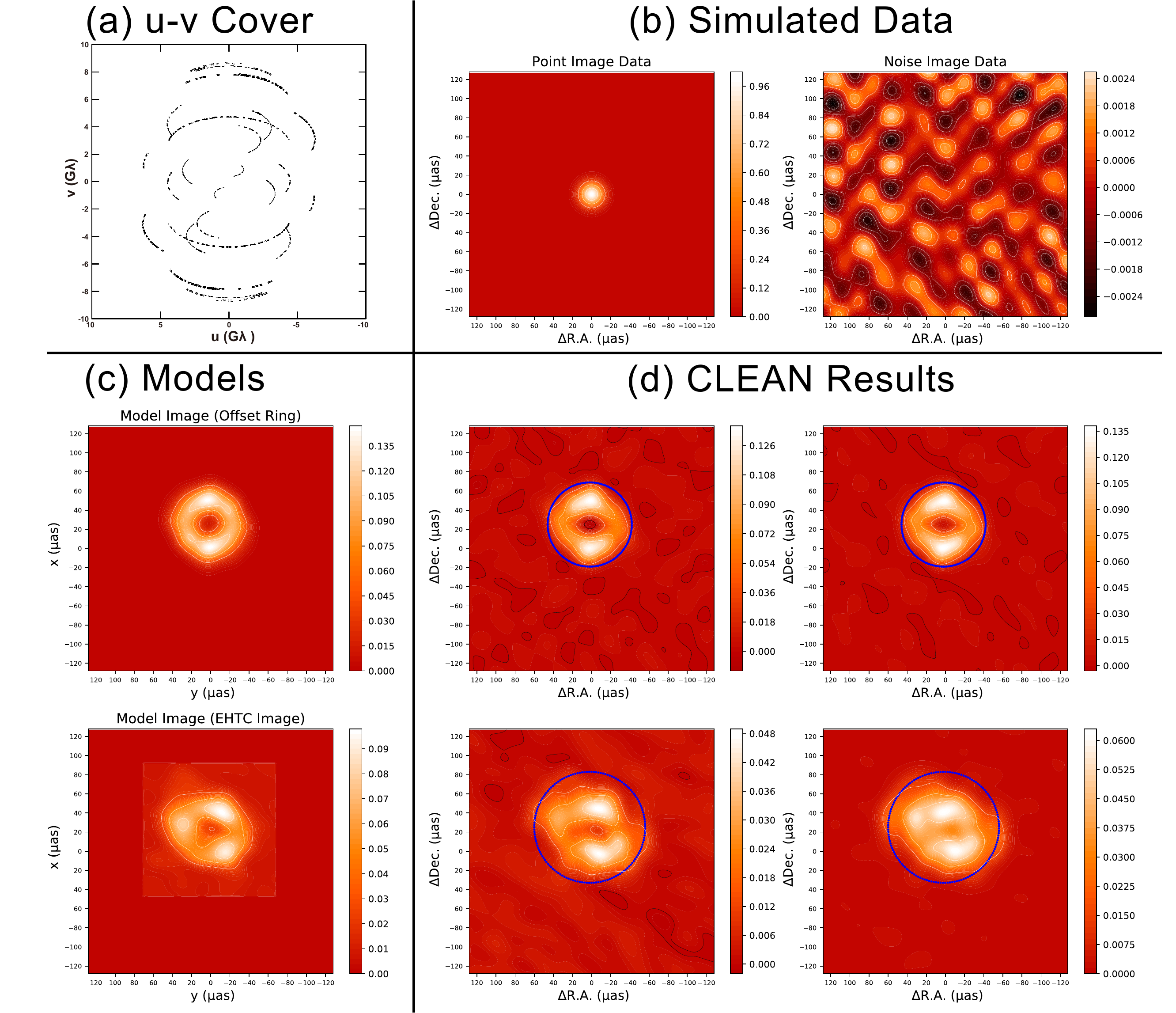}
\caption{
Resultant ring images from the simulated data (no ring) are shown. 
 Panel~(a) 
shows the practical \UVC coverage of the EHT array in 2017 for \SGR~observations. 
Panel~(b) shows images from the simulated data. The left shows the image of simulated point source data obtained using normal CLEAN ($\mathrm{20~\mu \rm as}$ beam convolution).
The right shows that of the simulated noise data obtained by Fourier transform of the data (dirty map).
Panel~(c) shows the model ring images used in self-calibration.
The upper shows the ring image with a diameter of $\mathrm{51.8~\mu \rm as}$ and a width of $\mathrm{20~\mu \rm as}$, which is obtained by CLEAN from the simulated visibility.
 The lower shows the EHTC ring image.
 Panel~(d) shows the CLEAN results obtained from the calibrated data sets by the self-calibration solutions. 
The blue circles in the panels indicate the BOX area specified by the field-of-CLEAN restriction. The unit of intensity for all images is Jy/BEAM.
}
\label{ringsim}
\end{figure*}

We performed self-calibration using the ring-like image models: a ring image with a diameter of $\mathrm{51.8~\mu \rm as}$ and the EHTC ring image. The solutions were used to "calibrate" the simulated data sets, and CLEAN imagings were performed with the CLEAN area limited by the BOX technique.
Figure~\ref{ringsim} shows the results of the simulations. 
 We were able to reproduce ring images similar to those used as image models in the self-calibration from both "no source structure with noise" and  "point without noise" data\footnote{Interestingly, the dirty map of "no source structure with noise" data shows a bumpy structure with a typical scale of about $\mathrm{50~\mu \rm as}$.}.
For the offset ring, the image is reproduced almost identically to the given model. For the EHTC ring, the resulting image is similar to the EHTC ring, but is not a perfect reproduction. 
This is probably due to the difference in data weighting.
The EHTC used their own data weighting to obtain their ring image (which is discussed in Section~\ref{Sec:DISC}), while our simulations do not use their weighting. 
Considering this point, it appears reasonable to suggest that our simulation results have essentially reproduced the equivalent of the EHTC ring image from non-ring data.
 Both "no source structure with noise" and "point without noise" data  are "calibrated" by self-calibration using a non-original image (ring) and imaging with a very narrow BOX setting, which produces a ring shape instead of the original image. This result is one of important indications that the EHT 2017 \UVC coverage is insufficient to accurately reproduce the observed object's structure and is likely to yield results consistent with the assumed image.
Obviously, because the self-calibration yields antenna-based calibration solutions, applying them to the data does not change the closure quantity. If we compare the closure quantity of the "calibrated" data with that of the visibility data Fourier transformed from the obtained image, they are not the same. It is clear that the ring-like images obtained here are artifacts.

\begin{figure*}
\includegraphics[width=0.95\textwidth,height=0.9\textheight,trim={0.25cm 0.75cm 0.75cm 0cm}, clip]{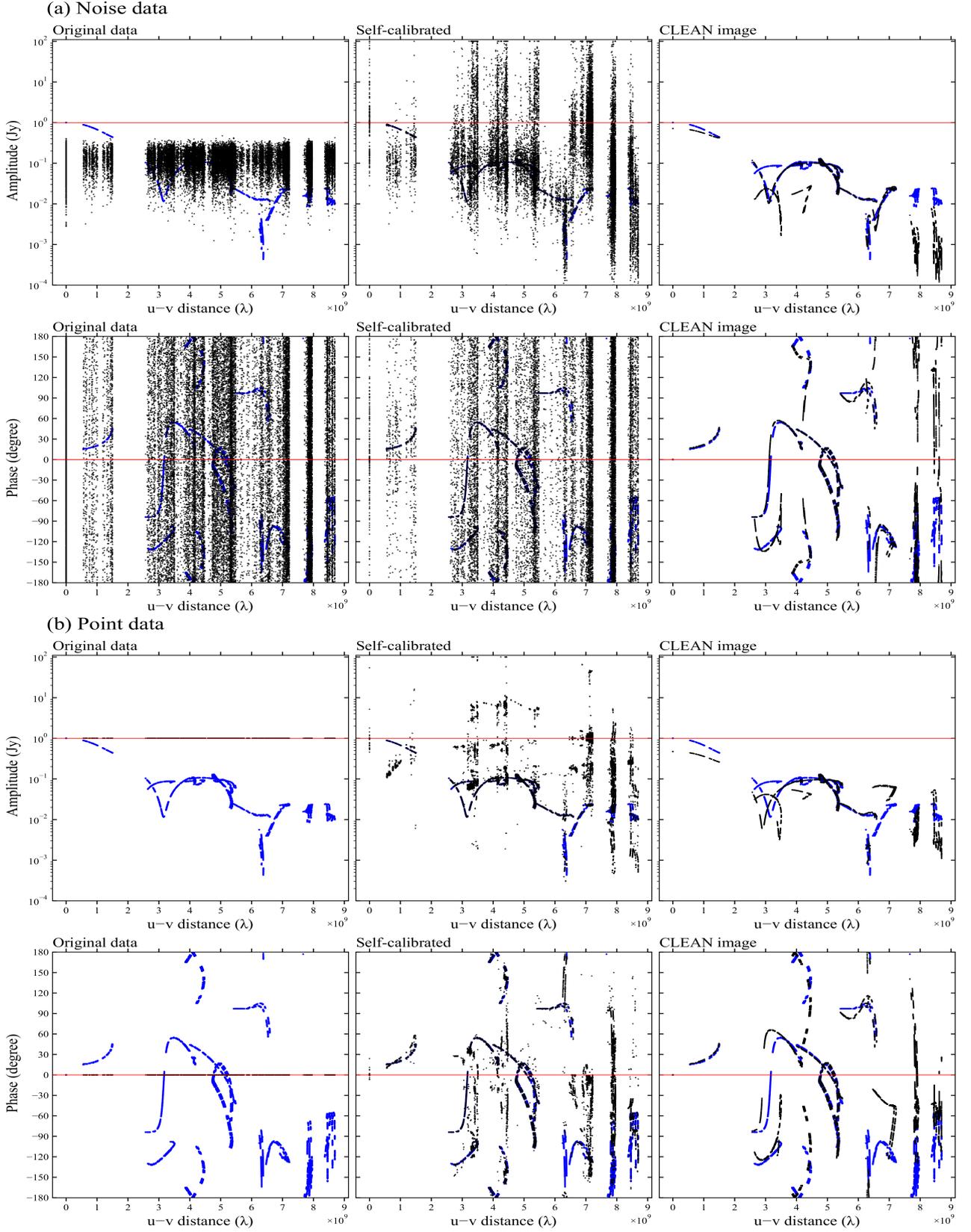}
\caption{
Strong impact of self-calibration and BOX setting on sparse \UVC data: 
 Panel~(a)  is for noise data (no structure with noise) and 
 Panel~(b)  is for point source without noise data. 
Their visibility distributions are indicated by black dots.
The red dots show the visibility distribution in the case of point source, and the blue dots show that in the case of the model image (EHTC ring).
The left shows the original visibility distributions~(amplitude and phase).
The central shows the visibility distributions after "corrected" by self-calibration with the ring model image.
The right shows the visibility distributions corresponding to the images obtained by CLEAN with the narrow BOX setting.
}
\label{ringsim0}
\end{figure*}

\vspace{\baselineskip}
From a different perspective, we examine the simulation results. Figure~\ref{ringsim0} presents the results in visibility space, rather than image space, and clearly demonstrates how self-calibration with a ring model and the narrow BOX setting in the CLEAN process significantly alter the originally non-ring visibility data, effectively reproducing a similar ring image.
The horizontal axis represents the \UVC distance, while the vertical axis shows the amplitude and phase. Red lines indicate the visibility distribution of a 1~Jy point source, and blue dots indicate those of a ring image. Simulated data points are shown in black.

In Figure~\ref{ringsim0}~(a), the visibility amplitude and phase in the case of "no source structure with noise" data are shown. 
The visibility amplitudes are Gaussian distributed around the mean corresponding to the noise intensity, and the phases are randomly and uniformly scattered, as shown in the left panels~(original data).
When the "no source structure with noise" data the are "calibrated" by the self-calibration solutions with the ring image model, as shown in the central panels~(self-calibrated), a significant number of data points match the visibility of the ring image (blue dots), while some do not.  
The right panels~(CLEAN image) show the visibility distribution of the resulting image from CLEAN processing with a narrow BOX setting on these "calibrated" data. Although not perfect, it closely corresponds to the visibility of the ring model.

In Figure~\ref{ringsim0}~(b), the visibility amplitude and phase in the case of "point without noise" data are shown.
All amplitudes are uniform at the given value of \textnormal{1~Jy}, and all phases are zero as shown in the left panels~(original data). 
~Incidentally, all closure phases are zero and all closure amplitudes are one.
 When the data are "calibrated" with self-calibration solutions with the ring-image model, as shown in the central panels~(self-calibrated), many, but not all, of the data points match the visibility of the ring-image (blue points). 
 The right panels~(CLEAN image) show the visibility distribution of the image obtained by CLEAN processing with a narrow BOX setting on these "calibrated" data. 
Although not perfect, it is also mostly consistent with the visibility of the ring model.

\begin{figure*}
\includegraphics[height=1.2\textwidth,trim=20 2 20 3,clip]{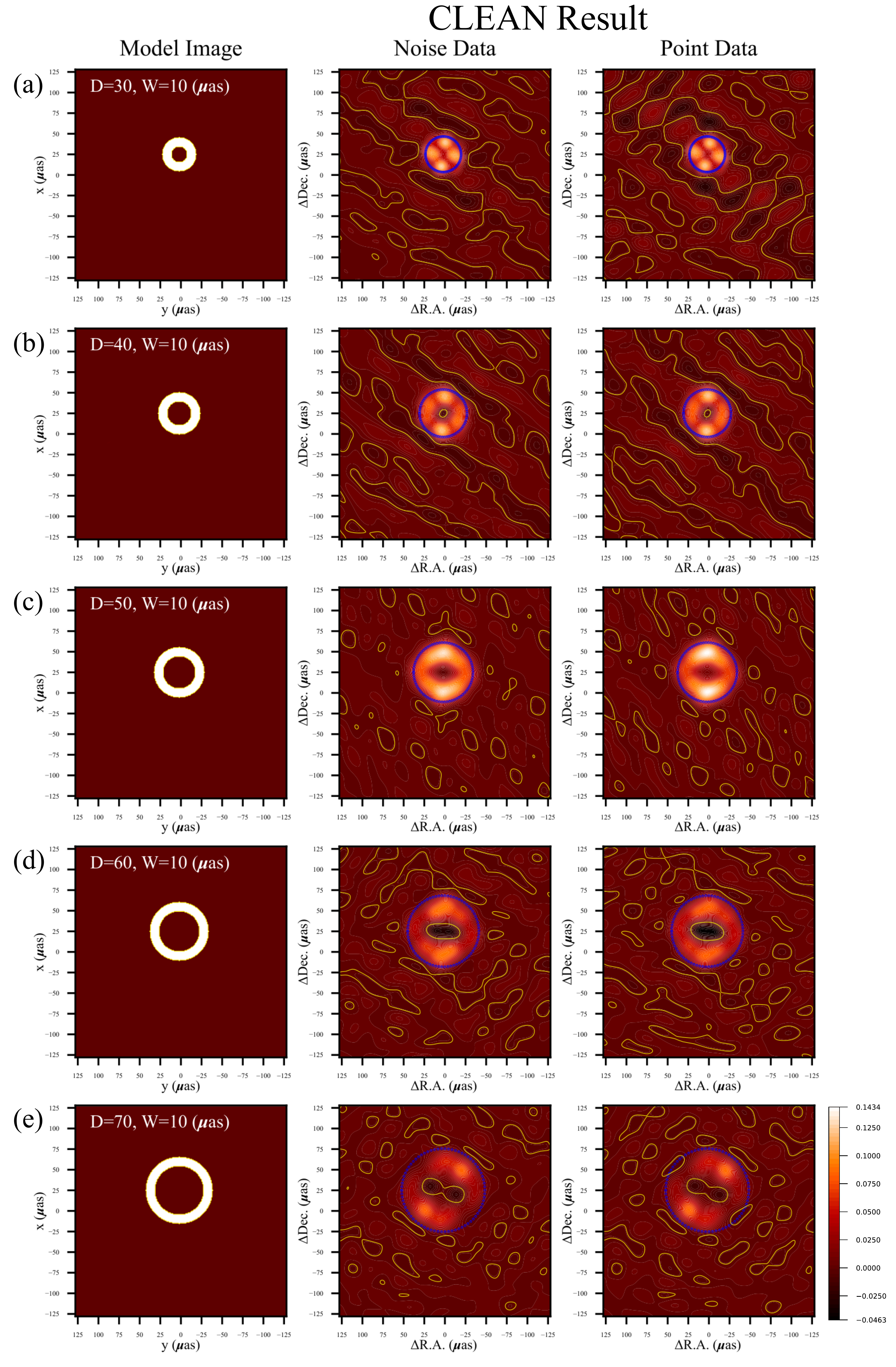}
\vspace{-3mm}
\caption{
Images resulting from changing ring diameters are shown. The model images used for self-calibration are shown in the left panels. The ring width is fixed at $\mathrm{10~\mu \rm as}$, and the total flux density is 1~Jy.
From (a) to (e), the diameters of the ring model are 
$\mathrm{30, 40, 50, 60}$, and~$\mathrm{70~\mu \rm as}$, respectively.
The resulting images obtained from simulated point source data are shown in the middle panels, and those obtained from simulated noise data are shown in the right panels. 
Contours indicate every \textnormal{10~\%} of the maximum brightness in all images. 
Solid lines indicate positive values and dashed lines indicate negative values. 
The blue circles in the panels indicate the BOX area specified by the field-of-CLEAN restriction. 
The restoring beam size is $\mathrm{20~\mu \rm as}$, and the intensity unit for all images is Jy/BEAM.
}
\label{ringsim2}
\end{figure*}

Furthermore, it is noteworthy that in the EHTC 2017 \UVC sampling data of \SGR~, rings with a diameter of $\mathrm{50~\mu \rm as}$ are most convincingly reproduced as artifacts, rather than rings with other diameters.
We ran the same simulations with different ring diameters to see if rings of different sizes could be regenerated.
We made rings with diameters ranging from $\mathrm{30}$ to $\mathrm{70~\mu \rm as}$, obtained self-calibration solutions using these ring image models, applied them to simulated visibility data sets for calibration, and performed CLEAN using the BOX technique to restrict the imaging area.
 The results of changing the ring diameter are shown in Figure~\ref{ringsim2}.
As before, the results show no significant difference between the two simulated data cases, suggesting that the PSF structure and imaging parameters (especially the box size) can affect the formed image regardless of the SNR of the data, whether infinite or infinitesimal.

\begin{figure*}
\includegraphics[width=\columnwidth,trim=7 8 7 7,clip]{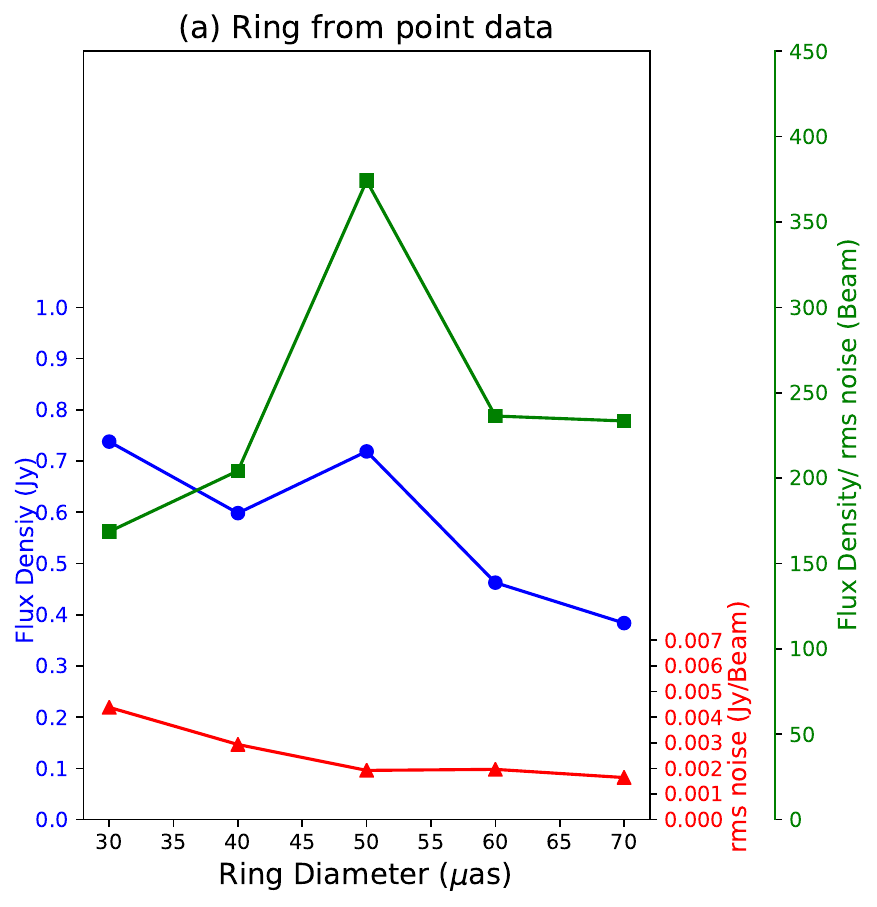}
\includegraphics[width=\columnwidth,trim=7 8 7 7,clip]{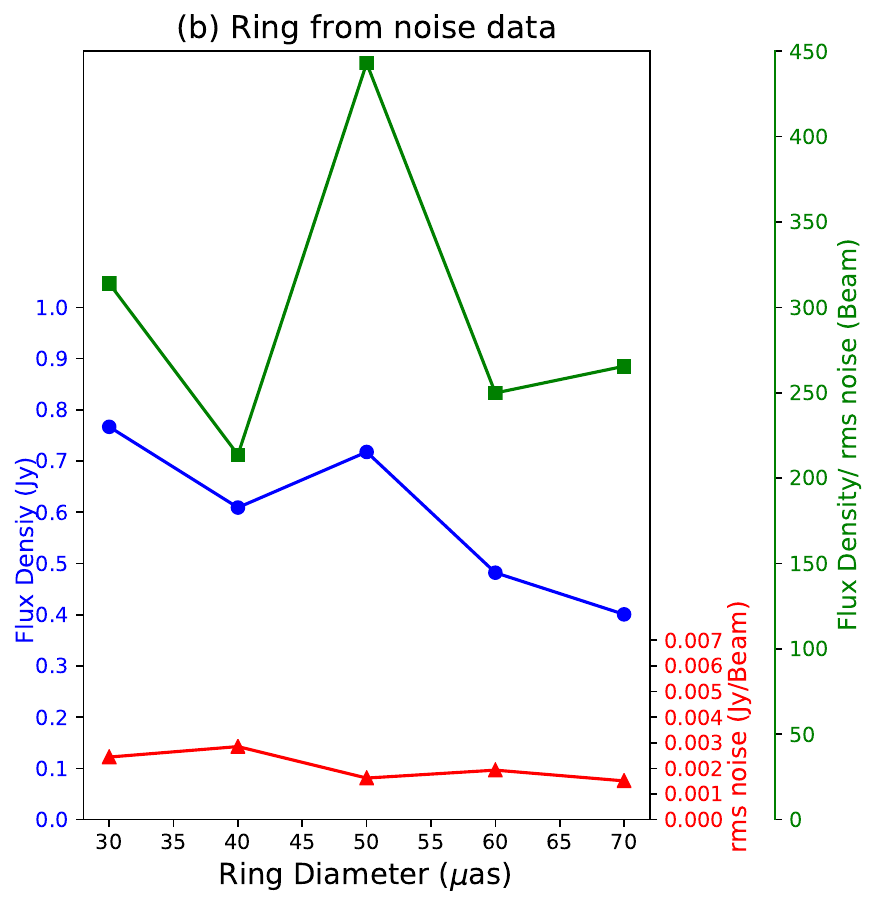}
\caption{
Properties of the images resulting from changing the ring diameter.
The flux densities (\textnormal{Jy}) summed from the CLEAN components are shown as blue plots.
Since the model images used for self-calibration are all \textnormal{1~Jy}, the closer the resulting flux density is to \textnormal{1~Jy}, the more similar the resulting image is to the model image.
The rms noise (\textnormal{Jy/beam}) measured in the region outside the ring image is shown as red plots; the smaller the rms noise, the more plausible the resulting image.
The green plots are the ratio of the flux density to the rms noise, which can be understood as another indicator of the plausibility of the resulting image. 
The higher the value, the more plausible the image.
In the cases from point simulated data are shown in Panel~(a) 
while those from noise simulated data are shown in Panel~(b).
}
\label{ringsim3}
\end{figure*}

The ring shapes seem to be well reproduced at $\mathrm{D=40, 50}$, and $\mathrm{60~\mu \rm as}$. 
In interferometric imaging synthesis, a negative brightness distribution often appears, but it is considered unnatural. 
In this simulation, the negative brightness distribution (surrounded by the yellow line in the panels) appears in all resulting images, but its area and depth vary. 
Comparing the size and depth of the negative areas, the image with $D=50~\mu \rm as$ shows the smallest negative area. 
Further quantitative comparison is attempted. 
Figure~\ref{ringsim3} shows the characteristics of the obtained ring images: the flux density summed from the CLEAN components, the rms noise of the image measured outside the ring, and the ratios of the flux density divided by the rms noise, which value is a measure of the plausibility of the obtained ring image. 
It can be seen that the ring image with a diameter of $\mathrm{50~\mu \rm as}$ shows the largest value and is the most plausible image among them.
Taken together, the ring formation is best reproduced in the case of $\mathrm{D=50~\mu \rm as}$.

These results indicate that in the imaging process with the \UVC coverage of the EHT array in 2017, the real image is not always correctly captured in the obtained results.
And $\mathrm{40\sim60~\mu \rm as}$-diameter ring images can be produced as artifacts, but especially among them, the ring shape with a diameter of $\mathrm{50~\mu \rm as}$ is the most likely to be produced as a most plausible artifact.

\subsection{Features that appear to be characteristic traces of the PSF structure in the EHTC simulated images.}\label{PSF-in-simulation}
The performance of CLEAN's deconvolution of the PSF structure is not perfect, and the resulting CLEAN map inevitably retains PSF effects. Other deconvolution  methods also do not completely remove PSF influences. 
The EHTC imaging methods would be no exception to this rule.
We show structures that appear to be influenced by the PSF structure are seen in the resulting images from the EHTC simulations shown in Figure 11 (a) of \cite{EHTC2022c}.
In EHT-imaging, the reproductions of the "Simple Disk" and "Elliptical Disk" show a shallow dip in the central regions of each disk, with a diameter of about $\mathrm{50~\mu\rm as}$. The reproductions of these disks in SMILI also show 
slight dips in their central regions. 
Also in SMILI, the $\mathrm{50~\mu\rm as}$ feature is more pronounced in the reproduction of the "Crescent (Ring)". 
Three sandbars appear from the central image towards the $\mathrm{PA=+45\DEG}$, $\mathrm{PA=0\DEG}$, and $\mathrm{PA=-45\DEG}$ directions, resulting in two inlets of about $\mathrm{50~\mu\rm as}$ in size. 
On the south side, an afterimage-like shape of the $\mathrm{50~\mu\rm as}$ ring from the model image can be observed. 
The THEMIS reproduction of the "Elliptical Disk" shows a ring structure extending in a north-south direction instead of the original disk image. The most interesting result of THEMIS is the reproduction of the "Simple Disk," which shows four small ring structures within the original disk area, each about $\mathrm{25~\mu \rm as}$ in diameter. These structures are likely due to the lack of $\mathrm{25~\mu \rm as}$ spatial Fourier components in the \UVC sampling (see Figure~\ref{Fig:uvs47}).

The foregoing indicates that the new EHTC imaging methods are not immune to the influence of the PSF, and that it is essential to examine the influence when evaluating the resulting images.
%
\subsection{The Case of the M\, 87 data in EHT 2017}\label{Sec:caseM87}
~For comparison with the \SGR case, we note here the EHT 2017 observations of \virgoa. From the data the EHTC reported a ring-like black hole shadow with a diameter of $42 \pm 3~\mu \rm as$ \citep{EHTC1,EHTC2,EHTC3,EHTC4,EHTC5,EHTC6}.
The diameter is certainly the size expected from the measured mass of the \virgoa SMBH ($\mathrm{6 \times 10^{9}~M_{\odot}}$~\citet{G2011}), while it is also consistent with the separation of $\mathrm{46~\mu \rm as}$ between the main beam and the first sidelobes in the PSF for \virgoa ~\citep{Miyoshi2022a}.
In terms of structure, the reanalysis found that there is a core-knots structure in the center from which a brightness distribution consistent with the previously observed jet structure extends westward~\citep{Miyoshi2022a, Miyoshi2022b, Miyoshi2024}. 
The comments below are mainly taken from ~\citet{Miyoshi2022a}.

%
\subsubsection{The sampled \UVC distribution}\label{M87uv}
\virgoa data were obtained in the same observing session as \SGR, but since \virgoa cannot be observed from the SPT in Antarctica, the number of stations is 7, one less than the \SGR observations, and 5 locations on the globe. 
The \UVC data lack samples corresponding to the EHT ring diameter; there is no sample at $\mathrm{d=44-46~\mu \rm as}$, and only very limited samples were observed just before and after this range.

\subsubsection{The PSF structure}\label{M87PSF}
The PSF structure for \virgoa ~is also remarkably bumpy. The main beam, which measures about $\mathrm{20~\mu\rm as}$~(FWHM). The first sidelobes, very close together, about $\mathrm{46~\mu\rm as}$ to the north and south, each reaching \textnormal{70~\%} of the height of the main beam. There is a negative minima deeper than \textnormal{-60~\%} of the height of the main beam at the midpoint between the main beam and the respective first sidelobe.
Like the case of \SGR, the size of the area including the main beam, the first sidelobe (one of the two) and the negative minima at the midpoint is the same as that of the EHTC \virgoa ring.
\subsubsection{Ring images from simulated data}\label{M87simring}
Also like the case of \SGR, the EHTC ring structures of \virgoa can be created from the simulated visibility data of a point image and a noise image.
The ring is very sensitive to the actual FOV size, i.e., the restriction by the BOX. If the FOV size is larger than $\mathrm{\sim 100~\mu \rm as}$ in diameter, the ring image cannot be formed well. 
The EHTC noticed this phenomenon, but treated it as a parameter setting to improve image clarity~\citep{EHTC4}.
We found that the most plausible ring image is produced when the center of the BOX is placed at the midpoint between the main beam and the first north sidelobe of the PSF, i.e. at the point of the deepest local negative minimum of the PSF.
%
\subsubsection{Indications of insufficient PSF deconvolution in the EHTC simulations }\label{M87simPSFR}
In the EHTC large-scale simulations for \virgoa data, the influence of the PSF structure is also observed: Figure 10 in \cite{EHTC4} shows that even for the double image model, the results from the two methods SMILI and eht-imaging include a faint ring structure of $\mathrm{\sim 40~\mu \rm as}$ diameter.
Note that the structure of the $\mathrm{\sim 40~\mu \rm as}$ interval is also present in the \virgoa image of ~\citet{Miyoshi2022a}'s independent analysis, although it is not conspicuous. 
In other words, the \virgoa ~\UVC data have a property of easily producing spurious structures at $\mathrm{\sim 40~\mu \rm as}$ intervals.

\subsubsection{Imaging pipelines using DIFMAP}\label{M87DIFMAP}
The EHTC opened the pipelines they used for the \virgoa imaging. 
Our findings about their imaging pipeline using DIFMAP are as follows.
In their procedure, a very narrow BOX with a diameter of $\mathrm{60~\mu \rm as}$, positioned not at the phase center but offset by $\mathrm{+22~\mu\rm as}$ to north was used for FOV restriction.
This BOX, covering the main beam and the first sidelobe (north) while excluding the second and further sidelobes, closely resembles the shape of the EHTC ring of \virgoa approximately $\mathrm{40~\mu\rm as}$. 
 In our view, such a narrow and offset BOX setting, if consistently applied, could significantly enhance the PSF structure effect, leading to the emergence of the EHTC ring as an intensified substructure within the PSF.
As evidence, when their pipeline was run without their BOX setting, it did not produce a ring-like image.

Our another concern is the repeated use of self-calibration to obtain both amplitude and phase solutions in most iterations during the hybrid mapping process. 
This could lead to an artifact in the final image.
\subsubsection{The residuals of the normalized amplitudes}\label{M87RNA}
As compared to the \SGR data,
the \virgoa data show significantly smaller normalized amplitude residuals, both for the EHTC ring and for the core-knots structure image of~\cite{Miyoshi2022a}.

\citet{Miyoshi2022a} shows those for an integration time of $\mathrm{t = 180~sec}$. 
For the first two days of \virgoa observations, the core-knots structure image of~\cite{Miyoshi2022a} shows the residuals of normalized amplitudes, 
$\mathrm{NR = 0.030 \pm 0.539}$, while the EHTC ring images show $\mathrm{NR = 0.148 \pm 0.933}$. For the last two days of observations, the image of~\cite{Miyoshi2022a} shows 
$\mathrm{NR = 0.127 \pm 1.259}$, while the EHTC ring images show $\mathrm{NR = 0.589 \pm 2.370}$. 
In contrast, those in the \SGR data show very large values:
$\mathrm{NR = 0.080 \pm 0.862}$ for our final image, and
$\mathrm{NR = 0.397 \pm 1.552}$ for the EHTC ring image.

\subsubsection{The residuals of the closure phases}\label{M87RCP}
The \virgoa data also have significantly smaller closure phase residuals for both the EHTC ring and the image of the core-knot structure by~\cite{Miyoshi2022a} than those in the case of the \SGR~data.
The core-knots structure also shows the same magnitude of closure phase residuals as those of the EHTC ring image.
The standard deviations of the closure phase residuals for a \textnormal{180 sec} integration are as follows: For the first two days of data, the core-knots image exhibits a standard deviation of 
$\mathrm{\sigma_{CK} = 40.5^\circ}$, compared to the EHTC image's~
$\mathrm{\sigma_{EHTC} = 38.5^\circ}$. For the data from the last two days, the core-knots image presents 
$\mathrm{\sigma_{CK} = 43.2^\circ}$, while the EHTC image shows $\mathrm{\sigma_{EHTC} = 43.7^\circ}$. 
Regarding the closure phase residuals, there appears to be no substantial difference between them.
Meanwhile, those in the \SGR data are $\mathrm{-0.68 \pm 58.08^\circ}$ for our final image, $\mathrm{-4.29 \pm 55.31^\circ}$ for the EHTC ring, which are larger residuals compared to those of \virgoa.

\subsubsection{Jet structure detection}\label{M87jet}
There is no mention of the famous jet of \virgoa~based on the EHT 2017 observations in the papers of~\cite{EHTC1,EHTC2,EHTC3,EHTC4,EHTC5,EHTC6}; \cite{Miyoshi2024} show that whether or not the ultrashort baselines in the data are used in the analysis determines the possibility of detecting the jet and faint structures around the core.

\section{Discussions}\label{Sec:DISC}
 Here, we provide a general discussion on the necessary steps for accurately imaging the black hole vicinity in \SGR~and \virgoa.
The EHT 2017 observations of \SGR and \virgoa~at \textnormal{230~GHz} were expected to reveal the structures in the vicinity of supermassive black holes.
Based on what we have shown so far, we think it is reasonable to conclude that the ring structures that are considered to be black hole shadows are derived from the PSF structure. As shown in Table~\ref{tab:ehtc_measurements}, the characteristic size of the PSF structure for each observation is consistent with the diameter of each ring.
The images from our independent analysis are not ring structures, and their consistency with the data comparable with or better than that of the respective ring images.
As for \SGR~, we identified an elongated, accretion disk like feature that is consistent with previous observations, but it is still possible that the image is blurred due to a large, time-varying effect similar to a subject blur.
As for \virgoa, the data reanalysis shows that a core-knots structure is still prominent in the center, the surroundings of the black hole are not in the light.
If that is the case, we are just at the beginning of our journey of reliable imaging studies in the vicinity of black holes.

~As we have shown, the PSF structure of the EHT in 2017 is very bumpy, and the influence of the PSF structure appears in the imaging results.
The PSFs of EHT in 2017 are also slightly different due to the different \UVC coverage for each of \SGR and \virgoa. The spacing between the main beam and the first sidelobe in the PSF is about $\mathrm{50~\mu \rm as}$ and $\mathrm{46~\mu \rm as}$, respectively. Both observational data tend to create artifact ring images with diameters corresponding to their respective spacings.

To obtain a PSF with a sharper beam, it is necessary to increase the \UVC coverage by obtaining a sufficient number of observing stations.
The number of antennas required to obtain reliable images of the vicinity of black holes in \SGR~without imaging assumptions has been suggested to be around 10~\citep{Miyoshi:04,Miyoshi:07}.
However, this study was performed at times when \SGR is believed to have no short-term variability. While this number of stations ensures that the \virgoa imaging study will be in progress, a larger number of stations is needed to accurately track the short-term variability of \SGR.
The short-term time variability of the \SGR has only recently been clearly identified.
The ALMA observations detected that the intensity of \SGR~change significantly during the VLBI observation period~\citep{Iwata2020, Miyoshi2019}.
It is reasonable to assume that the structure of \SGR~also shows variations on the same timescale.
As the EHTC also recognizes this fundamental issue~\citep{EHTC2022c}, 
this violates a fundamental condition required for interferometric imaging to be valid, i.e., the structure of the object must remain constant during the observation.
This situation is the same as the case of subject blur in photography, where the captured image is distorted by the subject's movement during exposure.
While mild variations could yield an acceptable approximation of the intrinsic  structure of \SGR,
it raises the question of whether the EHT array in 2017 captured the correct image in such situations.
We reviewed the EHTC's methodology used for mitigating~\SGR~variability. EHTC's approach involves using a variability noise modeling method that allows for the creation of static images even from time-variable data. The method adds variability noise to the uncertainty of every data point to mitigate the variation in the structure of \SGR. EHTC claims that "this parameterized variability noise model is generic and can explain well a wide range of source evolution, including complicated physical GRMHD simulations of \SGR~\citep{EHTC2022c}".
The variance of the additional noise budget is not a function of observing time, but depends on the \UVC length of the data points normalized by $\mathrm{4~\rm G\lambda}$ or additional parameter $\mathrm{u_{0}~\rm G\lambda}$, as shown in Equation (2) of~\cite{EHTC2022c}.
The specific normalization of the spatial Fourier components implies that the angular size of the object is already assumed before imaging the observational data.
As these normalization parameters are based on GRMHD simulation results~\citep{Georgiev:22}, then the physics of the object is also assumed.
In other words, this method assumes large restrictions on the physical properties and size of the object and attempts to mitigate the effects of variation in the observed object, it cannot be adapted to other variable objects.
 Results obtained by applying such methods are not direct images of observed sources, but rather shapes obtained by complex model fitting. 
Because the EHTC assumed the physics and the size of \SGR~to obtain its image, we think that their result on \SGR~does not allow one to search for observational facts about relativity and the physics of accretion disks.

The best way to obtain accurate images of time-varying objects is to 
 employ a sufficient number of antennas to ensure sufficient \UVC coverage within a time scale shorter than that of the object variation, allowing snapshots to be taken.
However, the number of  antennas currently available is  limited and snapshots are difficult to take.
The SMI method~\citep{Miyoshi:08} is designed to estimate the periodic variation component of the observed source structure. Employing this method, we are currently examining the potential to recover the time variability of \SGR~, and we anticipate presenting the findings in due course.
In the future, obtaining such a large number of antennas will be essential for reliable images from the \virgoa and \SGR~observations. 
However, for a reliable image of the \SGR~without the influence of short time variations, instantaneous and sufficient \UVC coverage for snapshot imaging is necessary. This means that 10 ground-based stations are not sufficient. A Low Earth Orbit (LEO) satellite for space VLBI~\citep{Asaki:09} that can fill the \UVC coverage in a short time would be more suitable.

We are now eagerly anticipating the upcoming observations with the expanded EHT array, which will improve the point spread function and provide clearer images of \virgoa and \SGR~that can be more easily understood.

\begin{table}
\centering
\begin{tabular}{@{}lcc@{}}
\toprule
 & \virgoa & \SGR \\
Predicted Shadow Size & 
$37.6^{+6.2}_{-3.5}$ or $21.3^{+5}_{-1.7}~\mu\text{as}$
& $\sim 50~\mu\text{as}$ \\
\midrule
\multicolumn{3}{@{}l}{EHTC Measurements} \\
\quad $D_{\text{ring}}$ & $42 \pm 3~\mu\text{as}$ & $51.8 \pm 2.3~\mu\text{as}$ \\
\quad $D_{\text{Shadow}}$ & - & $48.7 \pm 7.0~\mu\text{as}$ \\
\midrule
\multicolumn{3}{@{}l}{EHT PSF Structure} \\
\multicolumn{2}{@{}l}{\quad $1^{\text{st}}$ Sidelobe Position from the Main Beam} & \\
\quad & $46~\mu\text{as}$ & $49.09~\mu\text{as}$ \\
\multicolumn{2}{@{}l}{\quad $1^{\text{st}}$ Sidelobe Intensity Relative to the Main Beam} & \\
\quad & $+70~\%$ & $+49~\%$ \\
\multicolumn{2}{@{}l}{\quad Negative Minima at the Midpoint} & \\
\quad & $-60~\%$ & $-78.1~\%$ \\
\midrule
\multicolumn{3}{@{}l}{Restoring Beam Shape} \\
\multicolumn{3}{@{}l}{~~Default} \\
\quad ~~FWHM$_{maj\times min}$& $25.4~\times17.4~\mu\text{as}$ & $23.0\times15.3~\mu\text{as}$ \\
\quad ~~~Position Angle & $6.0\DEG$ & $66\DEG$ \\
~~EHTC Used &         &  \\
\quad ~~~FWHM & $20~\mu \text{as}$         & $20~\mu \text{as}$ \\
\bottomrule
\end{tabular}
\caption{Measurements of the EHTC rings and the characteristics of the corresponding PSFs.
Predicted shadow sizes, measured ring diameters and the restoring beam shapes are from EHTC papers.
Default beam values are from April 11 for \virgoa and April 7 for \SGR. 
The values of the PSF structures are from our measurements.
}
\label{tab:ehtc_measurements}
\end{table}
\section{Conclusion}\label{Sec:CR}
With  the conventional hybrid mapping, we reanalyzed the \SGR~data released by EHTC for the 2017 observations independently, and found a resulting image that differs from the one reported by EHTC. 
 Our analysis shows a roughly east-west elongated structure, which is consistent with previous millimeter wavelength VLBI observations.
The elongation is asymmetrical, with the east side being bright and the west side being dark.
We believe that our image is more reliable than the EHTC image because our image shows half the residuals of the EHTC image in normalized visibility amplitude although the residuals of the closure quantities of the two are comparable.
Assuming that the intensity ratio of bright and dark spots in the elongation is due to Doppler boost from the accretion disk rotation velocity, we estimate that we see the accretion disk at a radius of 2 to a few \RS ~from the center, rotating at \textnormal{60~\%} of the speed of light.
Given a central black hole mass of $\mathrm{\sim 4\times 10^6~M_{\odot}}$ and a distance of \textnormal{8~kpc}, we estimate the viewing angle of the rotating disk to be $\mathrm{\sim45\DEG}$.

While the EHTC analysis, based on calibrations with assumptions about the source's size and properties, selected the final image by prioritizing  appearance rate of the similar structure from a large imaging parameter space over data consistency.
The structure reported by EHTC is dominated by a bright, thick ring with a diameter of $\mathrm{51.8 \pm 2.3~\mu \rm as}$. 

In our view the ring-like image found by the EHTC is not the intrinsic structure of \SGR~but arises from the sparse \UVC coverage of the EHT array in 2017 i.e. to the corresponding $\mathrm{50~\mu \rm as}$-scale structure in the PSF.

The imaging using sparse \UVC data requires careful scrutiny of the PSF.
The estimated shadow diameter~($\mathrm{48.7\pm7~\mu \rm as}$) is equal to the spacing between the main beam and the first sidelobe of the PSF~($\mathrm{49.09~\mu \rm as}$), which immediately suggests a potential problem in the deconvolution of the PSF.
Also, this can be recognized from the fact that the ring image can be reproduced from simulated non-ring visibility data, and that a ring with a diameter of \textnormal{\FIMUS} is the most successfully produced.
 We found that internal inconsistencies in the closure quantities within the EHT 2017 \SGR~data, making it challenging to identify the most credible image from the amount of closure residuals. This issue is not related to the strong time variability of the \SGR. 
Investigating the cause of this phenomenon, including the correlation process of the data, is crucial for reliability of resultant images.

\section*{Acknowledgements}
We thank the anonymous referee for helpful comments that greatly improved the paper and the discussion.
This work is supported in part by the Grant-in-Aid from the Ministry of Education, Sports, Science and Technology (MEXT) of Japan, No.19K03939.
We thank the EHT Collaboration for releasing the network-calibrated \SGR~data.
The EHT2017 observations of \SGR~were performed with the following eight telescopes.
ALMA is a partnership of the European Southern Observatory (ESO; Europe, representing its member states), NSF, and National Institutes of Natural Sciences of Japan, together with National Research Council (Canada), Ministry of Science and Technology (MOST; Taiwan), Academia Sinica Institute of Astronomy and Astro- physics (ASIAA; Taiwan), and Korea Astronomy and Space Science Institute (KASI; Republic of Korea), in cooperation with the Republic of Chile. The Joint ALMA Observatory is operated by ESO, Associated Universities, Inc. (AUI)/NRAO, and the National Astronomical Observatory of Japan (NAOJ). 
The NRAO is a facility of the NSF operated under cooperative agreement by AUI. 
APEX is a collaboration between the Max- Planck-Institut für Radioastronomie (Germany), ESO, and the Onsala Space Observatory (Sweden). 
The SMA is a joint project between the SAO and ASIAA and is funded by the Smithsonian Institution and the Academia Sinica. 
The JCMT is operated by the East Asian Observatory on behalf of the NAOJ, ASIAA, and KASI, as well as the Ministry of Finance of China, Chinese Academy of Sciences, and the National Key R\&D Program (No. 2017YFA0402700) of China. Additional funding support for the JCMT is provided by the Science and Technologies Facility Council (UK) and participating uni- versities in the UK and Canada. 
The LMT project is a joint effort of the Instituto Nacional de Astr\.{o}fisica, \.{O}ptica, y Electr\.{o}nica (Mexico) and the University of Massachusetts at Amherst (USA). 
The IRAM 30-m telescope on Pico Veleta, Spain is operated by IRAM and supported by CNRS (Centre National de la Recherche Scientifique, France), MPG (Max-Planck-Gesellschaft,   Germany) and IGN (Instituto Geogr\.{a}fico Nacional, Spain).
The SMT is operated by the Arizona Radio Observatory, a part of the Steward Observatory of the University of Arizona, with financial support of operations from the State of Arizona and financial support for instrumentation development from the NSF. Partial SPT support is provided by the NSF Physics Frontier Center award (PHY-0114422) to the Kavli Institute of Cosmological Physics at the University of Chicago (USA), the Kavli Foundation, and the GBMF (GBMF-947). 
The SPT is supported by the National Science Foundation through grant PLR-1248097. Partial support is also provided by the NSF Physics Frontier Center grant PHY- 1125897 to the Kavli Institute of Cosmological Physics at the University of Chicago, the Kavli Foundation and the Gordon and Betty Moore Foundation grant GBMF 947.

\section*{Data availability}\label{Sec:DA}

Data available on request:
The original observational data are available on the EHTC website. 
The observational data and final solutions by self-calibration underlying this paper will be shared with corresponding authors upon academic request. We can provide the FITS data merging the two channels using  the AIPS task, DBCON. We will also include the final solution of our hybrid mapping there.


\clearpage
\begin{appendices}
%
\section{The normalized amplitude definition}\label{Sec:NAR-def}
The normalized amplitude definition is given below.

\begin{align}
\text{Residual of Normalized Amplitude} &= \nonumber \\
\nonumber \\
&\hspace{-12em} \frac{\text{Amplitude (image)} - \text{Amplitude (observed data)}}{\text{Amplitude (observed data)}}
\end{align}

The value is zero if the residual does not exist, or one if the amplitude indicated by the image is twice as large as that of the data. Thus, if the value exceeds 1, there is a very large difference in amplitude.

\section{The closure phase definition}\label{Sec:RCP-def}
The closure phase is a quantity that is immune to systematic errors resulting from antennas and depends solely on the brightness distribution of the source~\citep{Jennison:1958}. 
While the definition of closure phase is well-known in interferometric data analysis, it is provided here for clarity. The closure phase $\Phi_{\it{123}}$ for a triangle formed by three antennas, \textit{1}, \textit{2}, and \textit{3} is defined as follows.

\begin{equation}
\Phi_{\mathit{123}} \equiv \theta_{\mathit{12}}^{\text{obs}} + \theta_{\mathit{23}}^{\text{obs}} + \theta_{\mathit{31}}^{\text{obs}},
\end{equation}
where
\begin{align*}
\theta_{\mathit{12}}^{\text{obs}} &= \theta_{\mathit{12}} + (\phi_{\mathit{1}} - \phi_{\mathit{2}}), \\
\theta_{\mathit{23}}^{\text{obs}} &= \theta_{\mathit{23}} + (\phi_{\mathit{2}} - \phi_{\mathit{3}}), \\
\theta_{\mathit{31}}^{\text{obs}} &= \theta_{\mathit{31}} + (\phi_{\mathit{3}} - \phi_{\mathit{1}}).
\end{align*}

Here, $\theta_{ij}^{\text{obs}}$ is the observed fringe phase of the baseline between antennas \textit{i} and \textit{j}, 
$\phi_{i}$ is antenna based phase error, and $\theta_{ij}$ is the intrinsic phase due to the observed source structure.
If we substitute these in the equation,
\begin{eqnarray*}
 \Phi_{\mathit{123}} &=& \theta_{\mathit{12}}^{\text{obs}} + \theta_{\mathit{23}}^{\text{obs}} + \theta_{\mathit{31}}^{\text{obs}} \\
                     &=& \theta_{\mathit{12}} + (\phi_{\mathit{1}} - \phi_{\mathit{2}}) \\
                     &+& \theta_{\mathit{23}} + (\phi_{\mathit{2}} - \phi_{\mathit{3}}) \\
                     &+& \theta_{\mathit{31}} + (\phi_{\mathit{3}} - \phi_{\mathit{1}}) \\
                     &=& \theta_{\mathit{12}} + \theta_{\mathit{23}} + \theta_{\mathit{31}}.
\end{eqnarray*}

In the closure phase $\Phi_{\it{123}}$, antenna-based phase errors are cancelled, and the value of $\Phi_{\it{123}}$ is determined solely by the phases attributed to the structure of the observed source. 
The closure phase is zero with respect to the point source. 
Note that any baseline-based errors, if present, are not be cancelled.
Note also that thermal noise is not canceled.

If the true source image and the obtained image are identical, then the closure phase of the observed data and that of the visibility converted from the obtained image will have the same value, and the residual (the difference between them) will be zero. 
However, the reverse is not necessarily true in principle. A zero residual does not guarantee that the true source image has been obtained. On the other hand, if the residual is large, it is evident that the resulting image differs from the true image.\\

%
\section{The closure amplitude definition}\label{Sec:RCA-def}
Closure amplitude, like closure phase, is a quantity determined solely by the source structure and is free from antenna-based errors \citep{Felli:89,Readhead:80,Twiss:60}. 
Let us first consider the definition of closure amplitude.
The amplitude of visibility $\mathrm{|V_{ij}^{\text{obs}}|}$ observed by a baseline between antennas {\it i} and {\it j} can be expressed as the product of the gain errors $\mathrm{a_{i}}$,$\mathrm{a_{j}}$ due to the antennas and the amplitude $\mathrm{A_{ij}}$ due to the source structure.

\begin{equation}
\left| V_{ij}^{\text{obs}} \right| = a_{i} a_{j} A_{ij}.
\end{equation}

Suppose there are four antennas named 1, 2, 3, and 4.
Using the amplitudes of the observed visibilities through four baselines, namely $|V_{ij}^{\text{obs}}|$, where $\it{(i,j) = (1-2), (3-4), (1-3)},$ and $\it{(2-4)}$, we define the closure amplitude as follows:

\begin{equation}
\text{Closure Amplitude} \equiv \frac{|V_{\mathit{12}}^{\text{obs}}| \times |V_{\mathit{34}}^{\text{obs}}|}{|V_{\mathit{13}}^{\text{obs}}| \times |V_{\mathit{24}}^{\text{obs}}|} = \frac{A_{\mathit{12}} A_{\mathit{34}}}{A_{\mathit{13}} A_{\mathit{24}}}.
\end{equation}

This quantity is determined solely by the amplitude of the observed source structure. As the definition indicates, the visibility amplitudes of two baselines are divided by those of two other baselines. If an amplitude value with low SNR is in the denominator, the closure amplitude value can vary greatly due to its thermal noise. Therefore, it may give a value that is more unstable than that of the closure phase.
To stabilize the value of the closure amplitude, selecting only high SNR data for calculation would avoid the problem, but introduces another problem. In general, the higher the frequency of the spatial Fourier components, the lower the SNR. If closure amplitudes were calculated by omitting lower SNR data points, the resulting value would be influenced only by low spatial resolution components.
This would prevent us from accurately evaluating whether the resulting image captures the fine structures that correspond to the high spatial resolution

Note that any baseline-based errors, if present, will not be cancelled.
The closure amplitude is 1 with respect to the point source. 
If there is a baseline-based systematic error, it cannot be canceled, as is clear from the definition. Note also that thermal noise is not canceled.\\

Note that normalized closure amplitude is used in this paper for comparison. Its definition is similar to that of the normalized amplitude and can be shown as follows.

\begin{multline}
\text{Normalized Closure Amplitude} = \\
\\
\frac{\text{Closure Amplitude (image)} - \text{Closure Amplitude (observed data)}}{\text{Closure Amplitude (observed data)}}
\end{multline}

\section{Solutions of self-calibration in our analysis}\label{Sec:snplt}
\begin{figure}
\begin{center}
\includegraphics[width=\columnwidth,trim=0 1 0 0,clip]{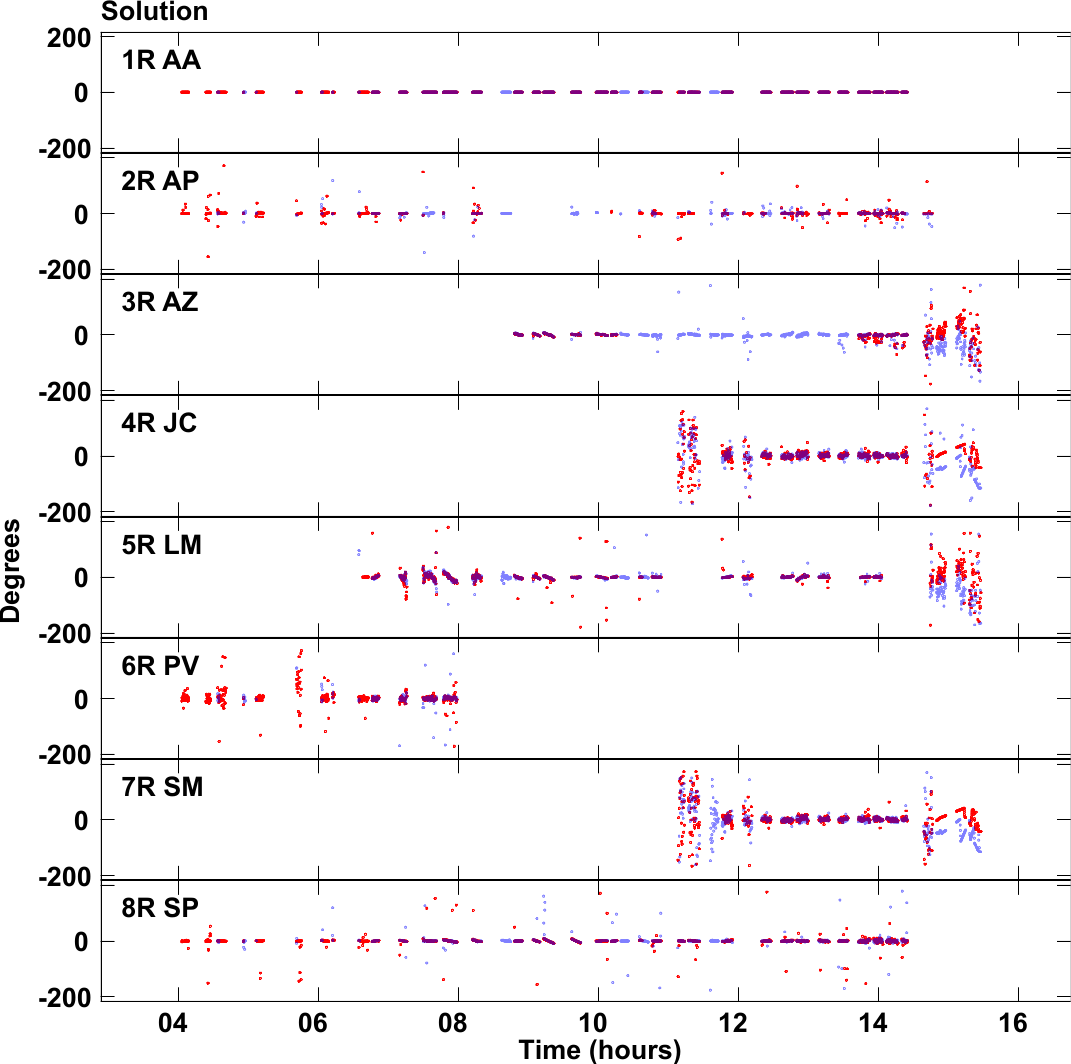}
\end{center}
\caption{
Initial phase-only self-calibration solutions obtained using one-point model for the data from low and high channels using CALIB in AIPS. 
The red dots represent the solutions for low channel data, while the blue dots represent the solutions for high channel data. 
}\label{Fig:SN1}
\end{figure}

In this section, we present the initial and final self-calibration solutions. 
Figure~\ref{Fig:SN1} shows the first self-calibration solutions, and Table~\ref{Tab:SN1} lists the self-calibration parameters used in the process. 
The first self-calibration solutions were obtained using one point model, and we found smaller differences between the two channels compared to those of the \virgoa public data. 

\begin{figure}
\begin{center}
\includegraphics[width=\columnwidth,trim=0 1 0 0,clip]{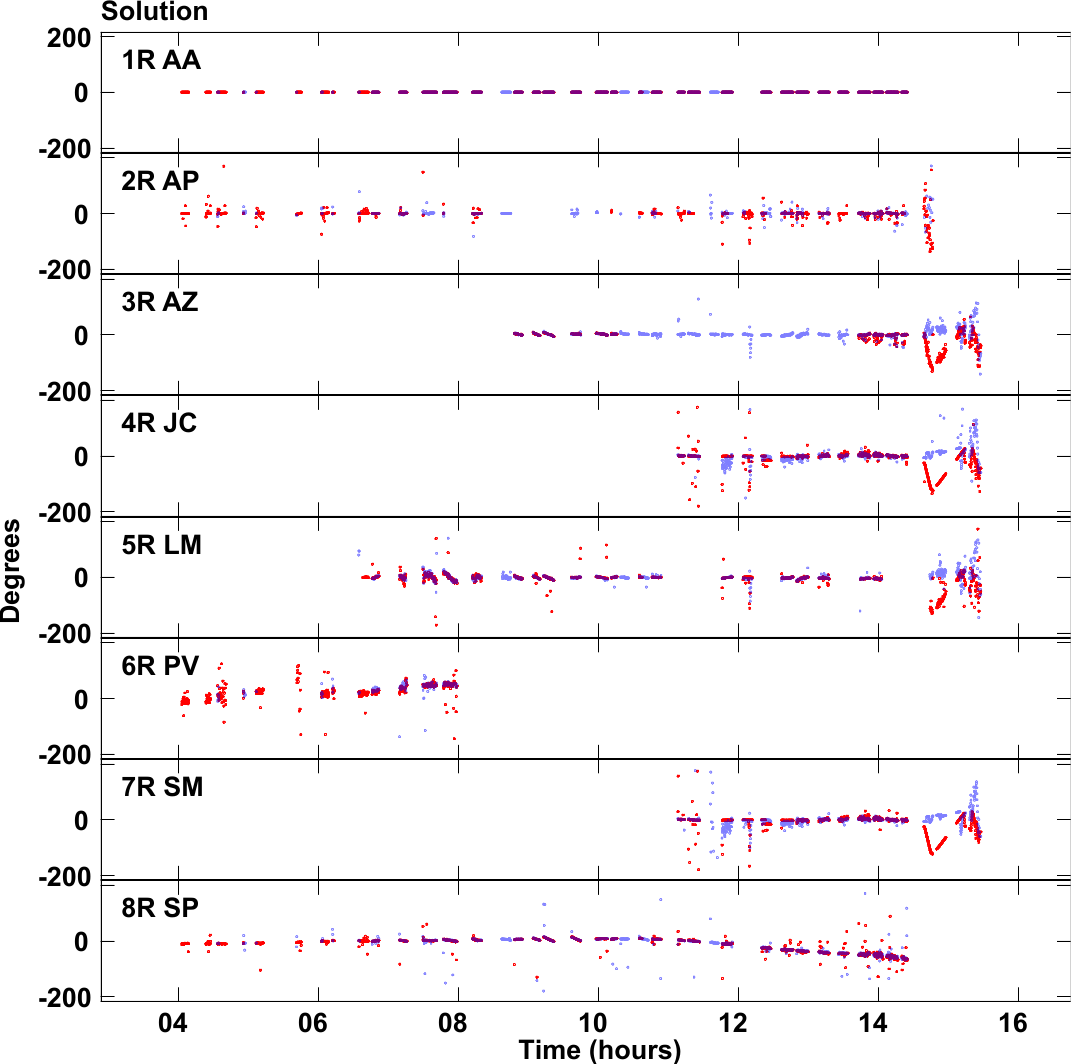}
\end{center}
\caption{
Final phase-only self-calibration solutions obtained for the data from low and high channels using CALIB in AIPS. The red dots represent the solutions for low channel data, while the blue dots represent the solutions for high channel data.
}\label{Fig:SN6}
\end{figure}

Figure~\ref{Fig:SN6} shows the final self-calibration solutions, where we kept the self-calibration parameters constant except for the image model.
The discrepancy between the first and final solutions is also minor. It is possible that a highly precise phase calibration was conducted during a-priori data calibration step (relative to the \virgoa public data), which could have contributed to these results.

\begin{table}
\begin{center}
\begin{tabular}{lr} \hline
Parameters        & \\ \hline \hline
 SOLTYPE &'L1'\\
 SOLMODE &'P' (phase only) \\
 SMODEL &1,0 (\textnormal{1~Jy} single point) \\
 REFANT &1 (ALMA)\\
 SOLINT (solution interval) &0.15 (min)\\
 APARM(1)&1\\
 APARM(7) (SNR cut off) &3\\ \hline
\end{tabular}
\end{center}
\caption{Parameters of CALIB for the first self-calibration.} \label{Tab:SN1}
\end{table}

\section{PSF (Dirty Beam) in the case of April 6, 2017 data}\label{Sec:uv46}
We present here the \UVC distribution and the corresponding PSF (dirty beam) structure of the April 6, 2017 observations. The imaging result of that day, as noted in the EHTC paper, showed that "although a ring feature appears in most of these reconstructions, it is less prominent."
In fact, the ring structure in the April 6 is not as clearly defined as that in the April 7. This may be due to the poor quality of the static image as \SGR~showed the large time variability on that day~\citep{EHTC2022c}.
In this section, we show that the difference in the robustness of the 
EHTC ring structure is less due to a structural time variation of \SGR~itself, but rather to the difference in the PSF structure between the April 6 and April 7 data.
The default restoring beam shape measured by EHTC is very similar for both observations.
    On April 6 it is $\mathrm{24.8 \times 15.3~\mu \rm as, PA= 67\DEG}$,
on April 7 it is $\mathrm{23.0 \times 15.3~\mu \rm as, PA= 66.6\DEG}$~\citep{EHTC2022c}.
However, a closer look at the structure of the PSFs shows significant differences between them.

\begin{figure}
\begin{center}
 \includegraphics[width=\columnwidth, trim=8 3 40 25, clip]{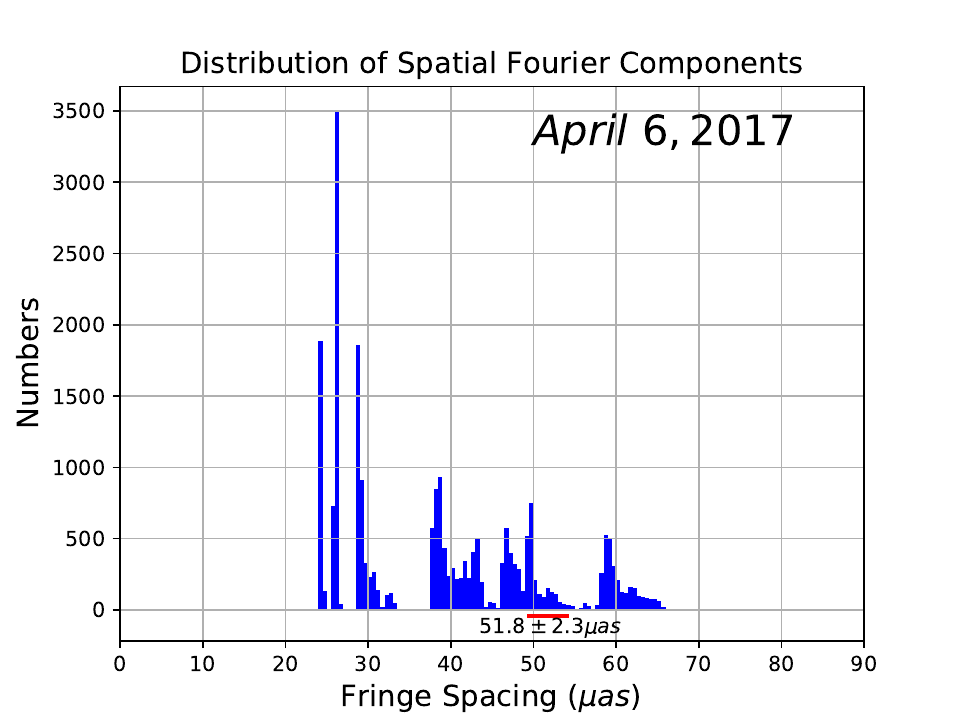}
\end{center}
\caption{
Distribution of the sampled visibility data from all baselines on April 6, 2017. The horizontal axis shows the fringe spacings of the data in units of $\mathrm{\mu\rm as}$. The vertical axis represents the number of sampled data. Spatial Fourier components ranging from approximately \textnormal{2.6 arcsec} to \textnormal{3.2 arcsec} are also present. 
Here, such large samples are omitted from the plot.
The red line segment displays the diameter of the ring as measured by the EHTC ($\mathrm{d = 51.8 \pm 2.3~\mu\rm as}$).
}\label{Fig:uvs46}
\end{figure}

 We show the \UVC sampling distribution for April 6, 2017 in Figure~\ref{Fig:uvs46}. The horizontal axis represents the size of the spatial Fourier component, and the vertical axis represents the number of samples.
The April 6 data have missing samples in several ranges, specifically 
$\mathrm{25\sim25.5~\mu \rm as}$, 
$\mathrm{27\sim28.5~\mu \rm as}$, and 
$\mathrm{34\sim37.5~\mu \rm as}$. 
While the missing ranges in the April 7 data are 
$\mathrm{24.5\sim25.5~\mu \rm as}$, and 
$\mathrm{27.0\sim28.5~\mu \rm as}$.
The April 6 data show a greater number of missing high frequency spatial Fourier components compared to the April 7 observations. However, the components corresponding to the ring size measured by EHTC ($\mathrm{d = 51.8 \pm 2.3~\mu\rm as}$) are still sampled, similar to the April 7 data.

\begin{figure*}
\includegraphics[width=1.0\textwidth,trim=55 65 50 75,clip]{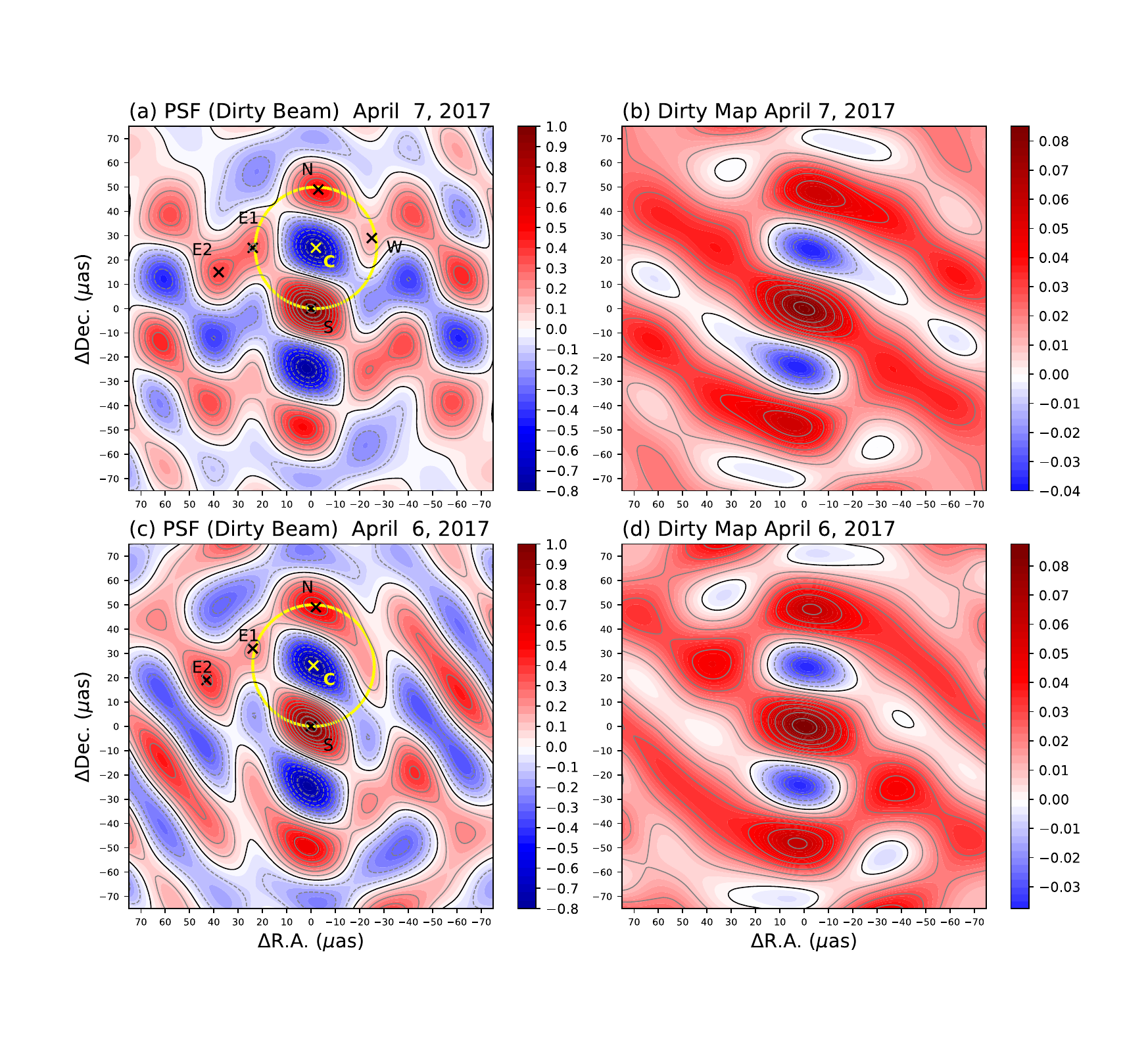}
\caption{
Point spread function (dirty beam) and dirty maps of the \SGR~observations.
Panels (a) and (b) at the top represent the observations from April 7, 2017,
while panels (c) and (d) at the bottom represent the observations from April 6, 2017.
The left panels (a) and (c) depict the point spread functions (dirty beam) of the two observations.
The right panels (b) and (d) show the dirty maps derived from Fourier transformations of the data calibrated by self-calibration solutions, with only phase calibration using a point source model.
}\label{Fig:psf4647}
\vspace{3cm}
\end{figure*}

The missing \UVC data impact the structure of the PSF. However, Figure~\ref{Fig:psf4647} shows that the most prominent bumpy structure in the corresponding PSF does not have the scales of the lacking spatial Fourier components but a scale of $\mathrm{\sim50~\mu \rm as}$ spacing, similar to the situation in the April 7, 2017 data.
This has important implications for the analysis of EHT data: the structure of the PSF cannot be inferred from the \UVC coverage plot alone, and it must be calculated in practice to fully understand its characteristics.
To avoid confusion about the quality of the \UVC data sampling, it is important to present the structure of the PSF instead of the \UVC coverage plot in scientific papers.

As discussed in Section~\ref{Sec:simData}, the $\mathrm{50~\mu \rm as}$ diameter ring can be formed from the PSF structure. 
The left panels, (a) and (c) in Figure~\ref{Fig:psf4647} show the PSF structures from two different days. Sidelobes comparable in height to the main beam are present, with deep negative minima existing between them. Among these, four peaks are located at approximately $\mathrm{25~\mu \rm as}$ from the northern deepest minimum, "C". These peaks are "E1", "N (one of the highest sidelobes)", "S (the main beam)", and "W". Together, they form an envelope that creates a $\mathrm{50~\mu \rm as}$ diameter ring. Presumably, the bright three spots observed on the EHTC ring correspond to the peaks of "E1", "N", and "S" on the envelope. These features are apparent in the PSF structure of the April 7, 2017 data.

While the PSF structure of the April 6, 2017 data is basically similar to that of the April 7 data, but there are differences in clarity of the four peaks. The clarity of these peaks in the PSF structure of the April 6 data is degraded compared to those in the April 7 data.

First, we discuss the "E1" peak, which is located in the eastern part of the envelope. It has an intensity of 0.305 relative to the main beam in the PSF structure of the April 7 data, but the intensity ratio in the April 6 PSF structure decreased to 0.224. Not only that, the "E2" peak, which is located farther east than the position of the "E1" peak, is actually brighter than the "E1" peak.
In the PSF structure of the April 7 data, the intensity ratio between "E1" and "E2" is 1.170, which means they have comparable brightness. However, in the PSF structure of the April 6 data, the ratio increases to 1.853, which means that the "E2" peak is about twice as bright as the "E1" peak. Therefore, another brighter structure is more likely to be created in the east of the $\mathrm{50~\mu \rm as}$ diameter envelope centered at C.
Then, the structure of the east side of the $\mathrm{50~\mu \rm as}$ diameter ring becomes very ambiguous. 

While peak "W" in the PSF is responsible for the western side of the EHTC ring, in the April 6 PSF, peak "W" disappears. As a result, the structure of the western side of the $\mathrm{50~\mu \rm as}$ diameter ring also becomes very ambiguous. Thus, the ring shape loses its east and west sides, making it more difficult to create the $\mathrm{50~\mu \rm as}$ diameter ring.

Admitting that the extreme time variability makes it difficult to obtain a static image of the object, we think that the reported changes in the EHTC ring structure between the two days are presumably due to the change in the PSF structure and not to that of \SGR.

\begin{table*}
\centering
\begin{tabular}{lcrcr}
\multicolumn{1}{l}{}&\multicolumn{2}{c}{7-Apr-2017}& \multicolumn{2}{c}{6-Apr-2017}\\ \hline 
Components            &Intensity&Position~~&{\rm Intensity}&{\rm Position}~~\\ 
                      &          &[$\mu \rm as]~~~$&&[$\mu \rm as$]~~~~\\ \hline
C~(deepest~minimum)    & -0.781&(~+2,~+26)&-0.778&(~+1,~+26)\\ 
E1                    & ~0.305&(+24,~+26)&~0.224&(+24,~+33)\\ 
E2~(out~of~the~circle)& ~0.357&(+38,~+16)&~0.415&(+43,~+20)\\ 
~~~~~{\it(ratio~of~E2/E1)} & {\it~1.170}&~~~~~~& {\it~1.853} &  \\  
N~(highest~sidelobe)  & ~0.491&~(~+3,~+50)&~0.556&(~+2,~+50)\\ 
S~(main~beam)         & ~1.000&~~(~~0,~~~~0)&~1.000&(~~0,~~~~0)\\ 
W                     & ~0.150&~~~(~-25,~+29)&{~~~~~~\it~disappeared} & \\ \hline
\end{tabular}
\caption{
Location and intensity of peaks around the deepest north minimum ”C” in the two observations, with intensity defined as the ratio of peak height to main beam height.
}\label{tab:my_label}
\end{table*}

\section{Difference of closure amplitude between the channels}\label{Sec:CA2}

Here we describe in detail the differences in closure amplitude between the channels noted in Section~\ref{Sec:RCA}.
In Figure~\ref{fig:res-closureA-sta}, not only the residuals of the normalized closure amplitudes, also we show the differences of those between the two recording channels. 

One of the differences between the recording channels, that of "low/high" (here the reference is the value of the high channel data), shows large values, which can be attributed to a lower signal to noise ratio of the high channel data. However, it cannot be explained why both the mean and the standard deviation of the channel differences become larger as the integration time increases.

A possibility is inferred that some baseline-based errors are present.
In our analyzed data, amplitude corrections have been applied to the baseline connecting the LMT station that are less than $2~G\lambda$ in projected length. 
It is possible that this acts as a closure amplitude breaker rather than a correction.
In general, it is difficult to imagine errors that originate from the baseline. 
If the correlation process is performed on each baseline and each recording channel, and each process performs delay and delay-rate tracking with individually different parameters, the closure amount would include the effect. 
It is also possible that some unknown error in the correlator system is at work.

We also examined the differences in closure amplitudes between channels for all publicly available data~(Appendix~\ref{Sec:betweeIFs}).
Again, significant differences were found in closures that should have been identical.
The fact that two of the different calibration methods (HOP, CASA) resulted in their individual differences between channels suggests that some errors may have occurred in the data processing after the correlation process.
The "BEST" data sets, from which 100 minutes of data were extracted, have significantly smaller standard deviations in the channel differences of both closure phase and closure amplitude than those of the data with the full observation durations. 
This suggests that the differences accumulate over a long period of processing.
If the correlation process for part of the baseline is performed with tracking parameters based on incorrect station positions and the data are processed for a long time, the errors will accumulate and not be cancelled within one sidereal day. 
It is possible that such a processing error may occur, but, it is very challenging to pinpoint the cause by investigations solely on publicly available EHT data.

Obviously, there is a discrepancy in the data themselves (at least between recording channels). This can make it difficult to select the optimal image based on a smaller residual of closure quantity.
\section{Closure difference between the two recording channels}\label{Sec:betweeIFs}
\begin{figure*}
\begin{center} 
 \includegraphics[width=\columnwidth,trim=45 4 33 8,clip]{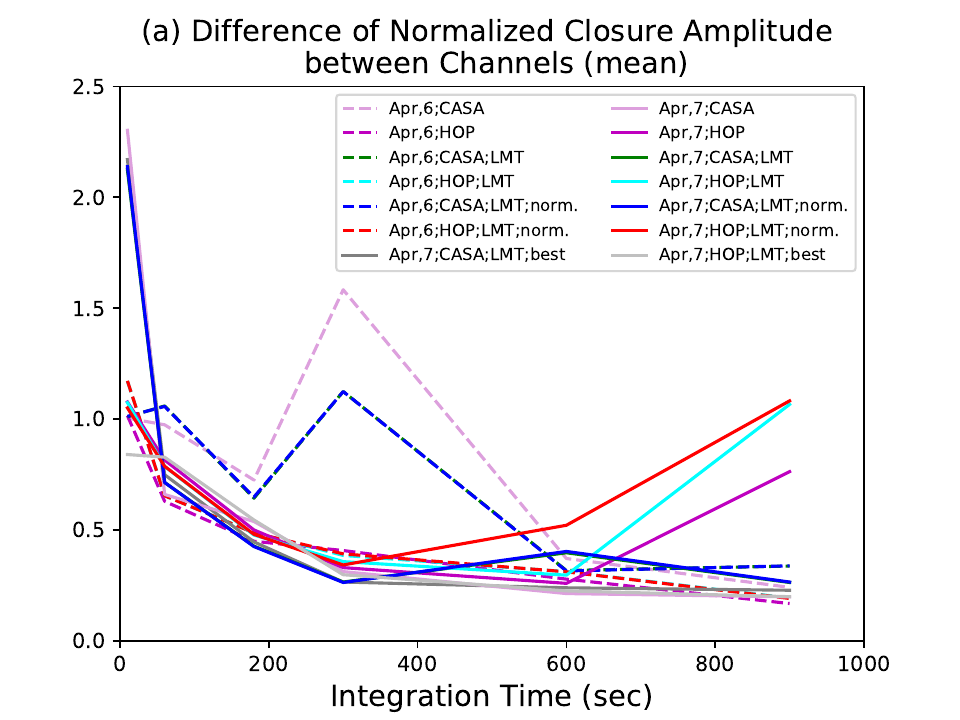}
 \includegraphics[width=\columnwidth,trim=45 4 33 8,clip]{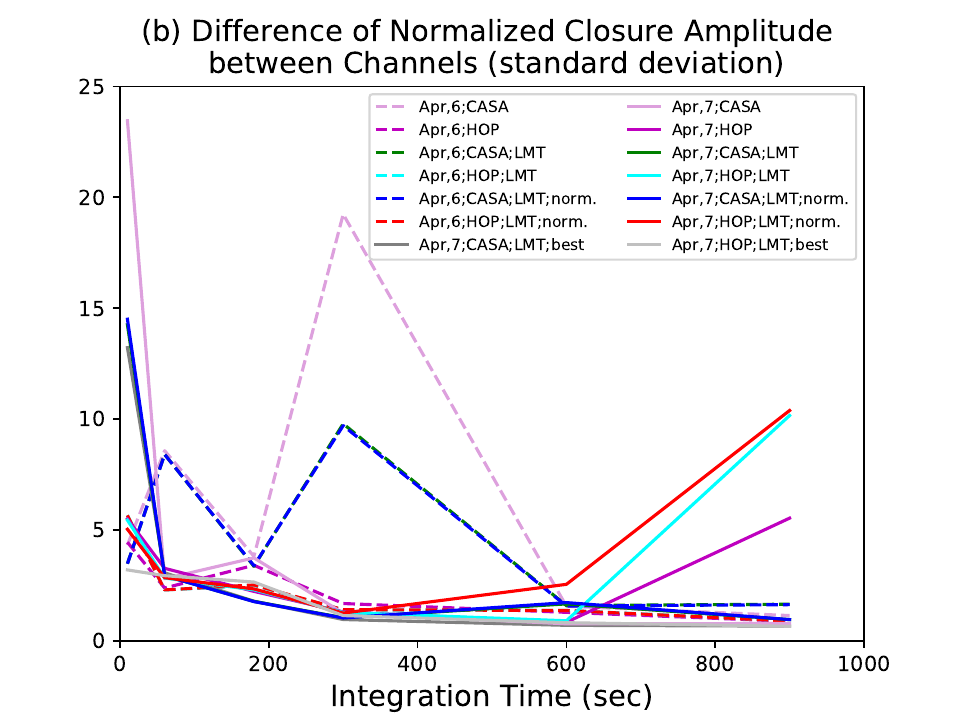}
\end{center} 
\caption{
Differences of closure amplitude between the two recording channels.
 Panel~(a)  
shows the means, and 
 Panel~(b)  
shows the standard deviations.
The solid lines indicate the values for the April 7 data, and the dashed lines indicate those for the April 6 data.
}\label{Fig:IFdif}
\end{figure*}

~In sections \ref{Sec:RCP} and \ref{Sec:RCA}, we compared the closure phase and closure amplitude of the observed data with those of our final image and the EHTC image. We found that both our final image and the EHTC ring image had rather large residuals, and increasing the integration time did not reduce these residuals. This is a phenomenon that is difficult to explain. After examining the closure differences between the data from the two recording channels, we concluded that the phenomenon was not caused by the reproduced images but rather by something in the data. To investigate this further, we decided to examine all of the public EHT data about \SGR.
In Table~\ref{Tab:IF-dif-CA}, we present the normalized difference in closure amplitude between the two recording channels, while in Table~\ref{Tab:IF-dif-CP}, we show the difference in closure phase between the two channels. Figure~\ref{Fig:IFdif} displays plots of both values.

Regarding the closure amplitude difference between the two channels, if the closure amplitude values of both channels are identical, the normalized closure amplitude difference should have a mean of zero and standard deviation of zero. If the values differ by a factor of 2, the normalized closure amplitude difference would be 1. 
As demonstrated in Table~\ref{Tab:IF-dif-CA}, the minimum mean is approximately 0.25, indicating that the closure amplitude in one channel is $25\%$ larger than that in the other channel. 
The minimum standard deviation is 0.66, indicating significant self-inconsistency in the closure amplitudes between the channels.
%

In terms of closure phase, the difference between the two channels is close to zero, as expected from theory. However, increasing the integration time does not bring the mean difference closer to zero. The standard deviation is also relatively large. If thermal noise is dominant, the variance should decrease by the $\mathrm{-0.5^{th}}$ power of the integration time T, but the actual standard deviation reduction is not significant.

As the definitions make clear, the closure quantity is determined solely by the source structure and is independent of antenna-based errors. Therefore, when two data sets are acquired simultaneously, their closure quantities should be identical. Even in cases where the structure of \SGR~exhibits very rapid time variations, there should be no difference in closure quantity between simultaneously recorded data sets. Hence, our findings here pertains not to the observed source, but rather to the quality of the data recording and a-priori calibration process.

We found that closure discrepancies occur in all publicly available EHT data sets of \SGR~observed in 2017, suggesting that the EHT data may not accurately represent the source structure.

If closure is not conserved, errors may arise that are not due to antenna-based factors but rather to individual baseline-based factors. For example, this can occur when correlating individual baselines using conflicting station positions and the Earth rotation parameters in the correlation process.\\

The following matters can be inferred from an examination of the details.
\vspace{-1.6em}
\begin{enumerate}
\item 
Comparing the data from day~6 and day~7, we do not see much change in the difference in closure quantity between the two channels.
Although the time variation of the \SGR intensity is more intense on day~6, the fact that it has little effect means that the difference in closure quantity between channels is not related to the short-term intensity variation of the source, but is caused by either the instruments~(presumably correlator) or the data processing.
\item
It appears that some baseline-based correction is applied not only in the correlation process, but also in the subsequent calibration process. Differences in closure amplitudes between channels are present in both processed data, but the degree of difference is greater in the CASA data than in the HOP data.
 Both the mean and the standard deviation of the closure amplitude differences between channels tend to be larger for the CASA processed data than for the HOP processed data. For the closure phase difference between channels, the standard deviation is the same, but the CASA-processed data tends to have a larger deviation from zero in the mean.
\item
The amplitude correction applied by EHTC to LMT is a baseline based correction, which in principle affects the closure amplitudes. However, there is no significant difference in closure amplitude between channels compared to data without such a correction. Rather, it seems to have the effect of reducing the standard deviation.
\item
Although the visibility amplitude is normalized by the light curve of \SGR from the single dish observations, it acts uniformly on all baseline data. Therefore, it should not affect the closure amplitude. There is no such effect on the difference in closure amplitude between channels.
\end{enumerate}
\begin{table*}
\small{
\begin{tabular}{lrrrrrr}
\hline
Normalized Closure Amplitude Difference&     &     &     &     &     &      \\   
~~~~~~~~Tint (sec)  &$10            $&$60             $&$180            $&$300            $&$600            $&$900 $\\ \hline
Apr,6;CASA          &$1.00\pm 4.32$&$0.97\pm 8.57 $&$0.73\pm 3.81 $&$1.58\pm 19.24$&$0.37\pm 1.56 $&$0.24\pm 1.14 $\\ 
Apr,6;HOP           &$1.02\pm 4.44$&$0.63\pm 2.38 $&$0.45\pm 3.41 $&$0.41\pm ~1.68 $&$0.28\pm 1.28 $&$0.17\pm 0.88 $\\ 
Apr,6;CASA;LMT      &$1.01\pm 3.48$&$1.06\pm 8.41 $&$0.64\pm 3.36 $&$1.12\pm ~9.79 $&$0.32\pm 1.59 $&$0.34\pm 1.65 $\\ 
Apr,6;HOP;LMT       &$1.17\pm 5.66$&$0.65\pm 2.29 $&$0.49\pm 2.50 $&$0.39\pm ~1.40 $&$0.31\pm 1.37 $&$0.19\pm 0.90 $\\ 
Apr,6;CASA;LMT;norm.&$1.01\pm 3.48$&$1.06\pm 8.40 $&$0.65\pm 3.39 $&$1.12\pm ~9.70 $&$0.32\pm 1.57 $&$0.34\pm 1.63 $\\ 
Apr,6;HOP;LMT;norm. &$1.17\pm 5.66$&$0.65\pm 2.29 $&$0.49\pm 2.52 $&$0.39\pm ~1.40 $&$0.31\pm 1.37 $&$0.19\pm 0.89 $\\ \hline
Apr,7;CASA          &$2.30\pm 23.45$&$0.66\pm 2.80 $&$0.54\pm 3.74 $&$0.31\pm 1.24 $&$0.21\pm 0.77 $&$0.20\pm ~0.78 $\\ 
Apr,7;HOP           &$1.08\pm ~5.50$&$0.82\pm 3.27 $&$0.50\pm 2.23 $&$0.33\pm 1.30 $&$0.26\pm 0.84 $&$0.76\pm ~5.53 $\\ 
Apr,7;CASA;LMT      &$2.12\pm 14.25$&$0.71\pm 2.94 $&$0.43\pm 1.78 $&$0.27\pm 1.04 $&$0.40\pm 1.66 $&$0.26\pm ~0.96 $\\ 
Apr,7;HOP;LMT       &$1.07\pm ~5.44$&$0.79\pm 2.83 $&$0.48\pm 2.30 $&$0.36\pm 1.28 $&$0.30\pm 0.90 $&$1.07\pm 10.16$\\ 
Apr,7;CASA;LMT;norm.&$2.14\pm 14.50$&$0.71\pm 2.94 $&$0.43\pm 1.77 $&$0.26\pm 1.03 $&$0.40\pm 1.72 $&$0.27\pm ~0.96 $\\ 
Apr,7;HOP;LMT;norm. &$1.05\pm ~5.03$&$0.79\pm 2.85 $&$0.48\pm 2.34 $&$0.34\pm 1.25 $&$0.52\pm 2.55 $&$1.08\pm 10.39$\\ 
Apr,7;CASA;LMT;best &$2.17\pm 13.22$&$0.75\pm 3.07 $&$0.45\pm 1.79 $&$0.27\pm 0.96 $&$0.24\pm 0.70 $&$0.23\pm ~0.66 $\\ 
Apr,7;HOP;LMT;best  &$0.84\pm ~3.20$&$0.83\pm 2.94 $&$0.55\pm 2.65 $&$0.30\pm 1.14 $&$0.23\pm 0.82 $&$0.20\pm ~0.66 $\\ \hline
\end{tabular}
}
\caption{
Difference in closure amplitude between the two recording channels (normalized).
The data names include the observation dates, recording bands, calibration pipeline names, and stages.
}\label{Tab:IF-dif-CA}
\end{table*}
\begin{table*}
\small{
\begin{tabular}{lrrrrrr}
\hline
Closure Phase Difference&     &     &     &     &     &      \\ 
Tint (sec)          &$10            $&$60             $&$180                  $&$300            $&$600            $&$900      $\\ \hline
Apr,6;CASA          &$-1.27\pm 81.14 $&$-0.20\pm 67.36 $&$-0.54\pm 58.38 $&$-0.16\pm 55.55$&$~0.99 \pm 49.21$&$-0.07 \pm 48.03$\\ 
Apr,6;HOP           &$-0.15\pm 80.48 $&$-0.49\pm 64.72 $&$~2.37\pm 57.31 $&$~0.09\pm 53.47$&$~0.70 \pm 46.30$&$-0.39 \pm 43.81$\\ 
Apr,6;CASA;LMT      &$-1.62\pm 81.53 $&$-0.30\pm 65.76 $&$-0.16\pm 59.68 $&$~0.20\pm 57.32$&$-0.55 \pm 52.15$&$-1.59 \pm 49.35$\\ 
Apr,6;HOP;LMT       &$-0.94\pm 80.12 $&$-0.45\pm 66.45 $&$~1.74\pm 58.77 $&$-0.79\pm 53.68$&$~0.18 \pm 48.25$&$-0.55 \pm 45.59$\\ 
Apr,6;CASA;LMT;norm.&$-1.62\pm 81.53 $&$-0.30\pm 65.76 $&$-0.10\pm 59.69 $&$~0.20\pm 57.32$&$-0.55 \pm 52.15$&$-1.57 \pm 49.36$\\ 
Apr,6;HOP;LMT;norm. &$-0.94\pm 80.12 $&$-0.45\pm 66.45 $&$~1.74\pm 58.75 $&$-0.79\pm 53.66$&$~0.18 \pm 48.25$&$-0.55 \pm 45.58$\\ \hline
Apr,7;CASA          &$-0.27\pm 82.95 $&$~0.90\pm 67.86 $&$~1.82\pm 55.62 $&$~2.96\pm 50.80$&$~0.74 \pm 48.22$&$~5.52 \pm 43.51$\\ 
Apr,7;HOP           &$~0.93\pm 83.00 $&$-1.29\pm 67.95 $&$~0.56\pm 54.31 $&$~0.64\pm 50.50$&$-0.99 \pm 48.27$&$~1.02 \pm 45.20$\\ 
Apr,7;CASA;LMT      &$-0.21\pm 83.08 $&$~0.02\pm 68.97 $&$~1.31\pm 55.94 $&$~3.31\pm 50.83$&$~1.89 \pm 50.41$&$~4.84 \pm 42.21$\\ 
Apr,7;HOP;LMT       &$~0.84\pm 83.37 $&$-0.24\pm 69.16 $&$~0.33\pm 54.89 $&$~1.30\pm 50.17$&$~0.77 \pm 48.76$&$~3.46 \pm 44.38$\\ 
Apr,7;CASA;LMT;norm.&$-0.21\pm 83.08 $&$-0.14\pm 68.96 $&$~1.29\pm 55.93 $&$~3.32\pm 50.82$&$~1.90 \pm 50.39$&$~4.90 \pm 42.20$\\ 
Apr,7;HOP;LMT;norm. &$~0.84\pm 83.37 $&$-0.24\pm 69.16 $&$~0.01\pm 54.88 $&$~0.85\pm 50.18$&$~0.21 \pm 48.79$&$~2.79 \pm 44.48$\\ 
Apr,7;CASA;LMT;best &$-1.28\pm 81.08 $&$~0.62\pm 64.95 $&$~0.14\pm 50.38 $&$~2.30\pm 37.37$&$~2.82 \pm 35.65$&$~1.25 \pm 26.66$\\ 
Apr,7;HOP;LMT;best  &$-0.06\pm 80.56 $&$-1.11\pm 63.46 $&$~0.21\pm 43.91 $&$~1.35\pm 33.80$&$~2.46 \pm 35.28$&$~1.20 \pm 24.62$\\ \hline
\end{tabular}
}
\caption{Difference in closure phase between the two recording channels in degree.
The data names include the observation dates, recording bands, calibration pipeline names, and stages.
}\label{Tab:IF-dif-CP}
\end{table*}

\begin{figure*}
\begin{center} 
 \includegraphics[width=\columnwidth,trim=45 4 33 8,clip]{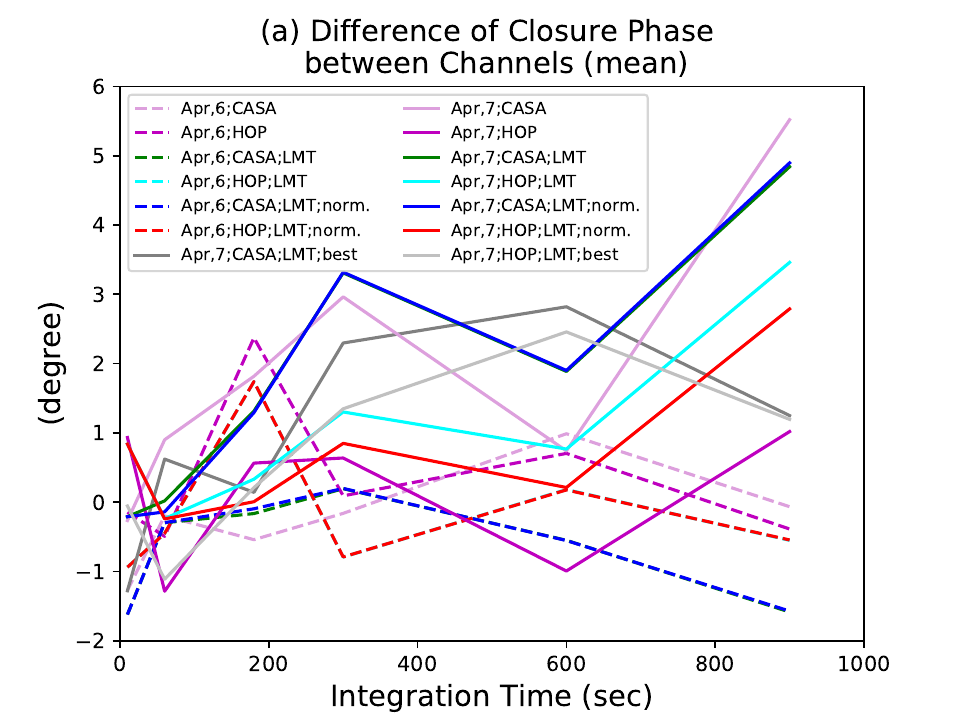}
 \includegraphics[width=\columnwidth,trim=45 4 33 8,clip]{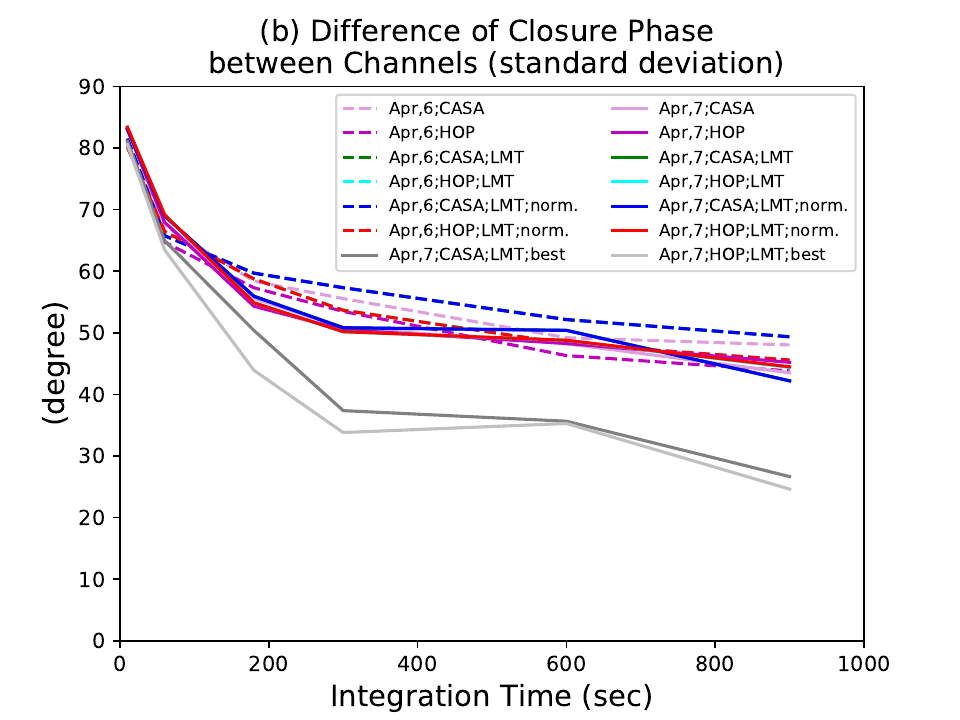}
\end{center} 
\caption{
Differences of closure phase between the two recording channels.
 Panel~(a)  
shows the means, and 
 Panel~(b)  
shows the standard deviations.
The solid lines indicate the values for the April 7 data, and the dashed lines indicate those for the April 6 data.
}\label{Fig:IFdif2}
\end{figure*}

\end{appendices}

\bsp	
\label{lastpage}
\end{document}